\documentclass{article}
\usepackage{hyperref}

\usepackage{authblk}
\usepackage{setspace}
\usepackage{amsmath,amssymb}
\usepackage{booktabs,xcolor,gensymb}
\usepackage[a4paper, margin=1in]{geometry}
\let\citep\cite
\let\citet\cite
\usepackage{bm,rotating}

\def\jobst#1{\textcolor{green}{ #1}}

\title{Degrees of individual and groupwise backward and forward responsibility 
        in extensive-form games with ambiguity,
        and their application to social choice problems}
\author[1]{Jobst Heitzig}
\author[2,1]{Sarah Hiller}
\affil[1]{Potsdam Institute for Climate Impact Research, PO Box 60 12 03, 14412 Potsdam, Germany, \href{mailto:heitzig@pik-potsdam.de}{heitzig@pik-potsdam.de}}
\affil[2]{Free University Berlin, Institute for Mathematics, Arnimallee 3, 14195 Berlin, Germany, \href{mailto:sarah.hiller@fu-berlin.de}{sarah.hiller@fu-berlin.de}}
\date{This version \today}

\def\eqdef{:=}

\def\rho{\varrho}
\def\ge{\geqslant}

\def\le{\leqslant}

\def\ignore#1{}

\def\T{\mathcal{T}}
\def\R{\mathcal{R}}
\def\eps{\varepsilon}
\def\N2{\frac{N}{2}}
\def\cm{\checkmark}

\newtheorem{proposition}{Proposition}
\newtheorem{conjecture}{Conjecture}

\begin{document}
\onehalfspacing

\maketitle
\begin{abstract} 
    Many real-world situations of ethical relevance, in particular those of large-scale social choice such as mitigating climate change, involve not only many agents whose decisions interact in complicated ways, but also various forms of uncertainty, including quantifiable risk and unquantifiable ambiguity. 
    In such problems, an assessment of individual and groupwise moral responsibility for ethically undesired outcomes or their responsibility to avoid such is challenging and prone to the risk of under- or overdetermination of responsibility.
    In contrast to existing approaches based on strict causation or certain deontic logics that focus on a binary classification of `responsible' vs `not responsible', we here present several different quantitative responsibility metrics that assess responsibility degrees in units of probability.
    For this, we use a framework based on an adapted version of extensive-form game trees and an axiomatic approach that specifies a number of potentially desirable properties of such metrics, and then test the developed candidate metrics by their application to a number of paradigmatic social choice situations.
    We find that while most properties one might desire of such responsibility metrics can be fulfilled by some variant, an optimal metric that clearly outperforms others has yet to be found.  
\end{abstract}

\section{Introduction}

The current climate crisis and its associated effects constitute one of the essential challenges for humanity and collective decision making in the upcoming years. An increase of greenhouse gas (GHG)\footnote{Prominently CO$_2$, but also methane, nitrous oxide and others.} concentrations in the atmosphere attributable to human activity leads to a warming of Earth's surface temperature by reducing the fraction of incoming solar radiation that is diffused back into space. An elevated mean earth surface temperature is however not a priori something reprehensible. Rather, it is the resultant effects that carry enormous dangers. Among these are the increased risk of extreme weather events such as storms and flooding, the rise of sea-levels or the immense losses of biodiversity, which have repercussions not only for the physical integrity of the planet but which pose direct threats to human life.\footnote{See for example \cite{gardiner2004} for a concise overview of the relevant climate science explained for non-climate scientists, or the IPCC and World Bank reports for more detail \cite{ipcc1.5spm,world2014turn}.}

Naturally, the public debate around this issue frequently invokes the question of \emph{responsibility}: Who carries how much backward-looking responsibility for the changes already inevitable, who is to blame; and who carries how much forward-looking responsibility to realise changes, who has to act?\footnote{What we call ``forward-looking'' or ex-ante responsibility 
is closely linked to the idea of 
obligation or duty, whereas what we call ``backward-looking'' or ex-post responsibility has also been called accountability, and relates to blame \citep{tamminga2019irreducibility,brahamvanhees2018}} 
As the following citation from Mike Huckabee, twice candidate in the US Republican presidential primaries, shows, the concepts of both backward and forward responsibility is used throughout the political spectrum: 
``Whether humans are responsible for the bulk of climate change is going
to be left to the scientists, but it's all of our responsibility to leave this
planet in better shape for the future generations than we found it.'' \cite{huckabee2007}




\paragraph{Existing work.} The existing body of work regarding this question can roughly be divided into two categories, via the perspective from which this question is addressed. On the one side there are considerations focusing on applicability in the climate change context, computing tangible responsibility scores for countries or federations, with the aim of shaping the actions being taken and a lesser focus on conceptual elegance and consistency \citep{botzen2008,mueller2009}. On the other side there is considerable work in formal ethics, aiming at understanding and formally representing the concept of responsibility in general with a special focus on rigour and well-foundedness, making it harder to account for messy real world scenarios (in realistic computation time) \citep{stit2001,brahamvanhees2012,chocklerhalpern2004,horty2001}.

It will be useful to highlight certain aspects of these works now. In the former set of works, and particularly also in public discourse, the degree of backward responsibility of a person, firm, or country for climate change is simply equated to cumulative past GHG emissions, or a slight variation of this measure \cite{guardian2011responsibility}. Certainly, this approach has one clear benefit, namely
that it is 
easy to compute on any scale, and also extremely easy to 
communicate
to a non-scientific audience. 
Similarly, certain authors assume a country's degree of forward responsibility to be proportional to population share, gross domestic product or some similar indicator, 
specifically in the debate about ``fair'' emissions allowances or caps \cite{ringius2002burden,poetter2019}.
However, unfortunately, such ad hoc measures violate certain properties that one would ask of a generalised responsibility account.\footnote{For example, using cumulative past emissions, population shares or GDP ratios all result in a strictly additive responsibility measure. If agent $i$ has a responsibility score of $R_i$ and agent $j$ one of $R_j$ the group consisting of agents $i$ and $j$ has a score of $R_i + R_j$. However, consider an example of two agents simultaneously shooting a third person. According to some intuitions, 
e.g., the legal theory of complicity \cite{kadish1985complicity}, they would then both be responsible to a degree larger than just half of the responsibility of a lone shooter. 
So we would need to either allow for group responsibility measures above 100\% (above total cumulative emissions/population share/GDP), or we would need to abandon additivity. Another issue of the cumulative emissions account is that many climate impacts are not directly proportional to emissions, a topic that will be discussed later on in this section.}

In the latter body of work, a principled approach is taken. Starting from considerations regarding the general nature of the concept of responsibility, formalisms are set up to represent these. These comprise causal models \cite{chocklerhalpern2004}, game-theoretical representations \cite{yazdanpanahdastani2016,brahamvanhees2012} or logics \cite{broersen2011mensrea,tamminga2019irreducibility}. A vast number of different aspects have been included in certain formalisations, such as degrees of causation or responsibility, relations between individuals and groups, or epistemic states of the agents to name but a few. Generally, these are discussed using reduced, well-defined example scenarios and thought experiments capturing certain complicating aspects of responsibility ascription. 

Additionally, there are investigations into the everyday understanding of the various meanings of the term `responsibility' \cite{vincent2011} as well as empirical studies regarding agents' responsibility judgements in certain scenarios, showing a number of asymmetry results \cite{nelkin2007}. However, we are here not concerned with mirroring agent's actual judgements, but rather with a normative account, so we will not go into detail about these.

\paragraph{Research question.} The present paper places itself in the category of a principled and formal approach, but aims at keeping in mind the practical applicability in complex scenarios.
Also, we want to relocate the space of discussion in the formal community by proposing a set of responsibility functions that, rather than cautiously distributing responsibility and tolerating under-determination (or {\em voids}), distribute responsibility somewhat more generously, evading certain forms of under-determination, but sometimes resulting in what might be seen as over-determination. The ``correct'' function is probably somewhere in between, and we think it is helpful to examine the space of possible solutions from several ends. It might be useful to add that our work is normative, not descriptive. We aim at representing ways in which responsibility {\em should be} ascribed, not the ways in which people in standard discussion generally {\em do} ascribe it or are psychologically inclined to perceive.  

We introduce a suitable framework that is able to represent all relevant aspects of a decision scenario. In some core aspects this is an extension of existing frameworks, in others we deviate from the previous work. Subsequently, we will suggest candidate functions for assigning real numbers as degrees of responsibility (forward- as well as backward looking) that have certain desirable properties. 

Deliberation regarding which climate abatement goal is to be reached but also who will contribute how much in the joint effort to mitigate climate change is often carried out in the political sphere, with various voting mechanisms in place. It is therefore particularly interesting to determine measures of responsibility when the deliberation procedure is given by a specific voting rule. We will address this question for a set of voting scenarios and our proposed responsibility functions.


\paragraph{Method.} We will follow an axiomatic method as it is used in social choice theory in order to enable a well-structured comparison between different candidates for responsibility functions \cite{axiomaticmethod}. That is, after determining a framework for the representation of multi-agent decision situations with ambiguity and corresponding responsibility functions as well as their properties, we begin by determining a set of simple, intuitive and basic properties that one might want a prospective candidate for an appropriate responsibility function to fulfil. 
Our framework is based on the known concept of \emph{extensive-form games}, with added features to represent the additional information, or rather lack thereof, that we want to include here.




\noindent
\paragraph{Specific aspects to be considered.} The above outline already shows several features of anthropogenic climate change that complicate responsibility assignments and occur in similar forms in other real-world multi-agent decision problems in which uncertainty and timing play a significant role. We will now highlight and discuss several features that our framework will need to include, as well as certain aspects that we treat differently from existing work. One important idea is to avoid allowing agents to refuse taking on responsibility by recurring to a certain calculation, even though according to some intuitions they do carry (higher) responsibility. We will suggestively call such an argumentation scheme \emph{dodging,} and the corresponding modelling aspect \emph{dodging-evasion}.

First of all, the effects of climate change are the result of an \emph{interaction} of many different actors: corporations, politicians, consumers, organisations, groups of these, etc.\ all play a role. Next, there is considerable \emph{uncertainty} regarding the impacts to be expected from a given amount of emissions or a given degree of global warming. While for some results we can assign probabilities and confidence intervals, for others this cannot be done in a well-founded way and beyond specifying the set of possible alternatives one cannot resolve the \emph{ambiguity} with the given state of scientific knowledge.

When several models give similar but slightly diverging predictions, for example, as is very often the case, we cannot assign probabilities to either of the models being `more right' than the others. What we can say however, is that each of the predictions is within the set of possible outcomes (given the premises, such as a certain future behaviour). The same goes for varying parameters within one and the same model.

Contrastingly, in a large body of work concerning effects of pollution, or warming, predictions are associated with a specified probability. Take for example the IPCC reports, such as the well known statement about the remaining carbon budget if warming is to be limited to 1.5 degrees: ``[\ldots] gives an estimate of the remaining carbon budget of [\ldots] 420 GtCO$_2$ for a 66\% probability [of limiting warming to 1.5\degree C above pre-industrial levels]'' \cite{ipcc1.5spm}. 
In many cases, both aspects of uncertainty --- ambiguity and probabilistic uncertainty --- are combined by speaking about intervals of probabilities, which in particular the IPCC does pervasively \cite{ipcc_uncertain}.

We argue that it is equally important to take note of the additional information in the probabilistic uncertainty case (often called `risk' in economics\footnote{Since we use the term `risky' in this article for a different concept, we stick to the term `probabilistic uncertainty' here.}) as of the lack thereof in the ambiguity case. It is known that the distinction between probabilistic and non-probabilistic uncertainty is important in decision making processes, and we want this to be reflected in our attribution of responsibility \citep{ellsberg1961}.

As a further particularity, the effects of global warming do not scale in a linear way with respect to emissions. 
With rising temperatures, so called `tipping elements' such as the Greenland or West Antarctic ice sheets risk being tipped \cite{lenton2019tippingpoints}: once a particular (but imprecisely known) temperature threshold is crossed, positive feedback leads to an irreversible procession of higher local temperatures and accelerated degradation of the element.\footnote{Ice reflects more sunlight than water. Thus, if a body of ice melts and turns into water, this will retain more heat than the ice did, leading to higher temperatures and faster melting of the remaining ice. As is stated in \citep{schellnhuberrahmstorf2016}: ``The keywords in this context are non-linearity and irreversibility''. Note that not all tipping elements are bodies of ice --- coral reefs or the Amazon rain forest also rank among them. The examples with corresponding explanation were chosen for their simplicity.} This initially local effect then aggravates global warming and may contribute to the tipping of further elements \cite{kriegler2009imprecise,wunderlingetal2019} 
adding up to the already immense direct impacts such as in case of these examples a sea level rise of several meters over the next centuries \citep{schellnhuberrahmstorf2016}.

We 
think 
that this nonlinearity should be reflected at least to some extent in the resulting responsibility attribution. This constitutes another argument for deviating from the --- linear --- cumulative past emissions accounts mentioned above \citep{botzen2008,mueller2009}.

In contrast to existing formalisations of moral responsibility in game-theoretic terminology \cite{brahamvanhees2018}, we include a temporal dimension in our representation of multi-agent decisions by making use of extensive-form game trees rather than normal-form games. This temporal component is also featured in formalisations using the branching-time frames of {\em stit}-logics.\footnote{Note that normal-form game-theoretical models correspond to a subclass of \emph{stit} models \cite{duijf2018}. Similarly, extensive-form games can also be represented as a \emph{stit}-logic \citep{broersen2009}, but we don't pursue this further here as the additional features that we will include would complicate a logical representation and this is not currently necessary to express what we want to.}

However, we do not take into account the temporal distance of an outcome to the individual decisions that led to it. Unlike in the ongoing debate in the environmental economics community regarding the discounting factors to be employed when considering future damages, with the prominent opposition between William Nordhaus and Nicholas Stern \citep{nordhaus2007b,stern2010} 
and its ``non-decision'' by a large expert panel led by Ken Arrow \cite{arrow2013should}, our account is not directly affected by any form of discounting. This is because while quantitative measures of welfare depend on notions of preferences, degrees of responsibility depend on notions of causation instead. Still, if the effects of an action disappear over time because of the underlying system dynamics (e.g., because pollutants eventually decay) and if this reduces the probability of causing harm much later, this fact can be reflected in the decision tree via probability nodes.

As another difference to existing formalisations we do not generally allow for assumptions regarding the likelihood of another agent's actions. We consider every agent to have free will, 
which we interpret to imply that while agents might have beliefs about others' behaviour, such beliefs cannot be seen as ``reasonable'' beliefs that provide justification in the sense of \cite{baron2016justification}. In other words, while beliefs about others' actions may influence the psychologically perceived degrees of responsibility of the agents, it should not influence a normative assessment of their responsibility by an ideal ethical observer or ``judge''.

Note that even \cite{brahamvanhees2012} state that the important feature with respect to judging other's actions in a tragedy of the commons application is that ``[b]efore the game was played, each agent assigned at least some positive probability to the strategy combination the others actually did play'', i.e.\ an unstructured set of possible outcomes suffices. 
If we want to express probabilistic uncertainty, this can be done, but without specifying an actor.

Unlike those accounts of responsibility focusing on the so-called `necessary element of a sufficient set' (NESS) test to represent causation, such as \cite{brahamvanhees2012}, we employ here the probability raising account as stated by \cite{vallentyne2008bruteluck}: ``I shall assume that the relevant causal connection is that the choice increases the objective chance that the outcome will occur --- where objective chances are understood as objective probabilities''. This account lends itself to our approach as we specifically want to discuss situations in which the outcome occurs with a given probability, and it enables a straightforward representation of {\em degrees} of responsibility. Note however, that unlike \cite{vallentyne2008bruteluck} we do not refer to agents' {\em beliefs} regarding the probabilities.

\paragraph{Paradigmatic examples and their evaluation.} In order to better understand the proposed frameworks as well as the responsibility functions, we will refer to a number of paradigmatic examples, mostly known from the literature or moral theory folklore, for illustration purposes. 
Like thought experiments in other branches of philosophy, such as the famous trolley problem, 
these examples have been selected because they each represent an interesting aspect of responsibility attribution in interactive scenarios with uncertainty that will come up later in the delineation of the proposed responsibility functions.

\begin{figure}
    {\bf (a)}\includegraphics[width=.4\textwidth]{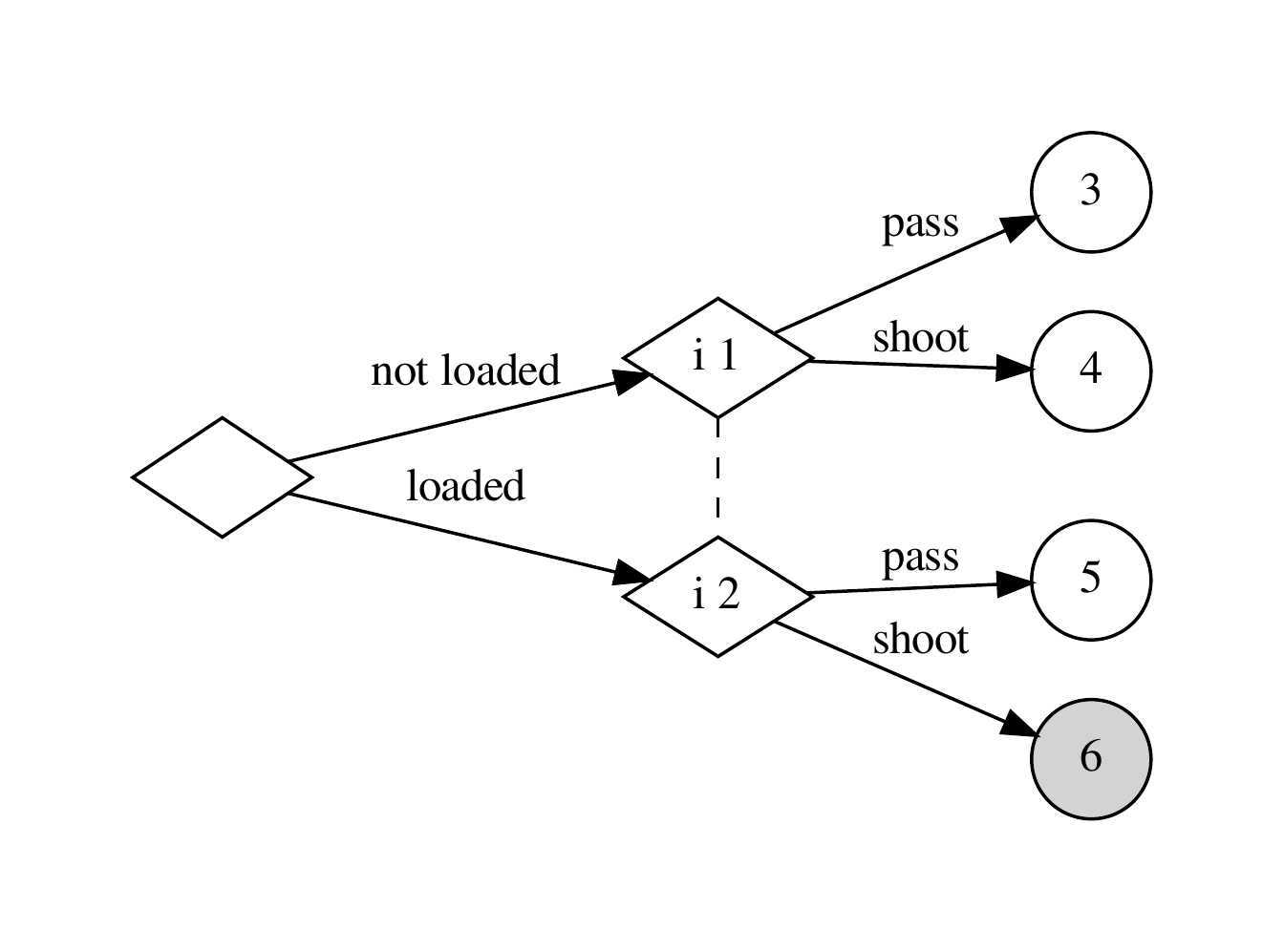}\quad
    {\bf (b)}\includegraphics[width=.4\textwidth]{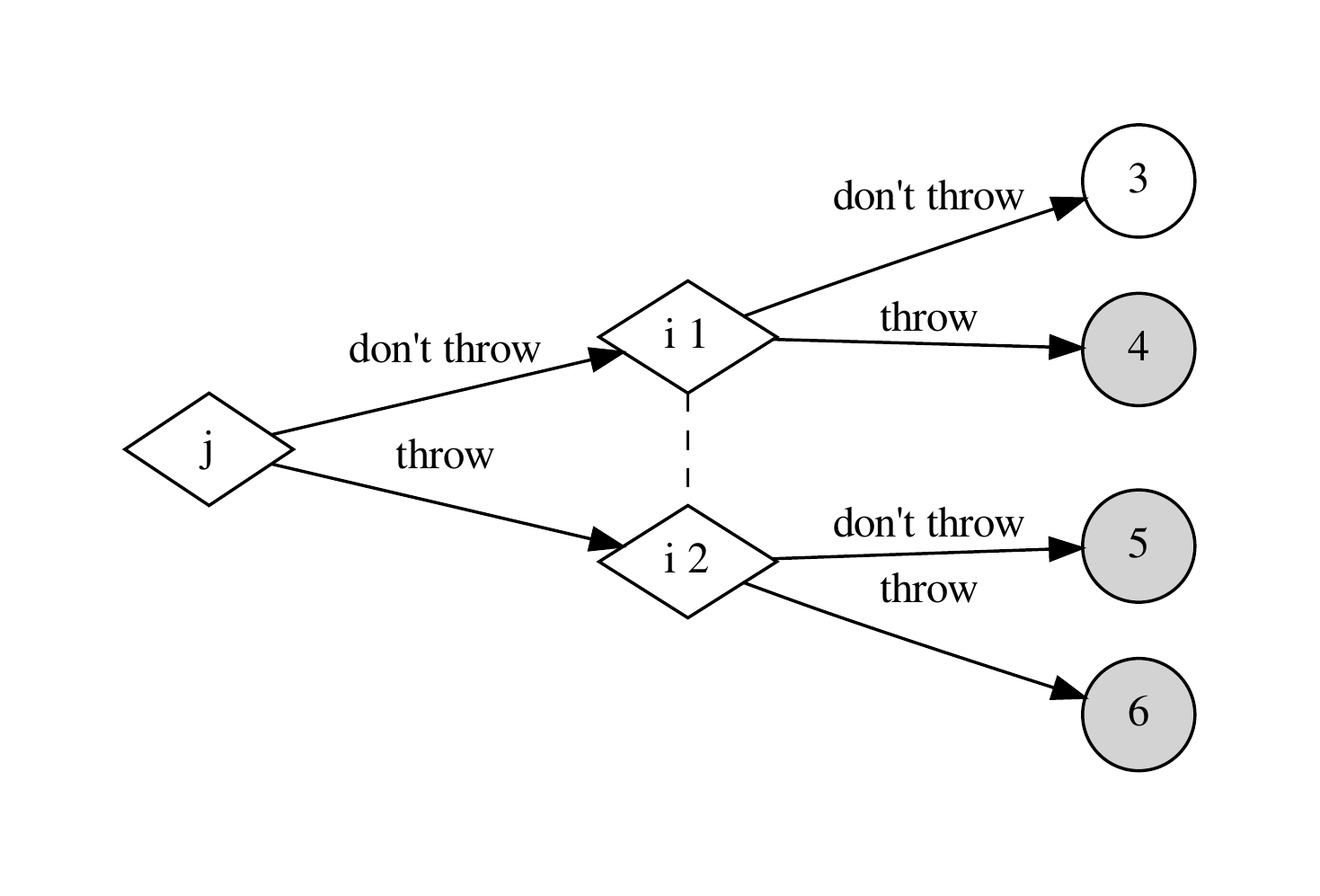} \\
    {\bf (c)}\includegraphics[width=.35\textwidth]{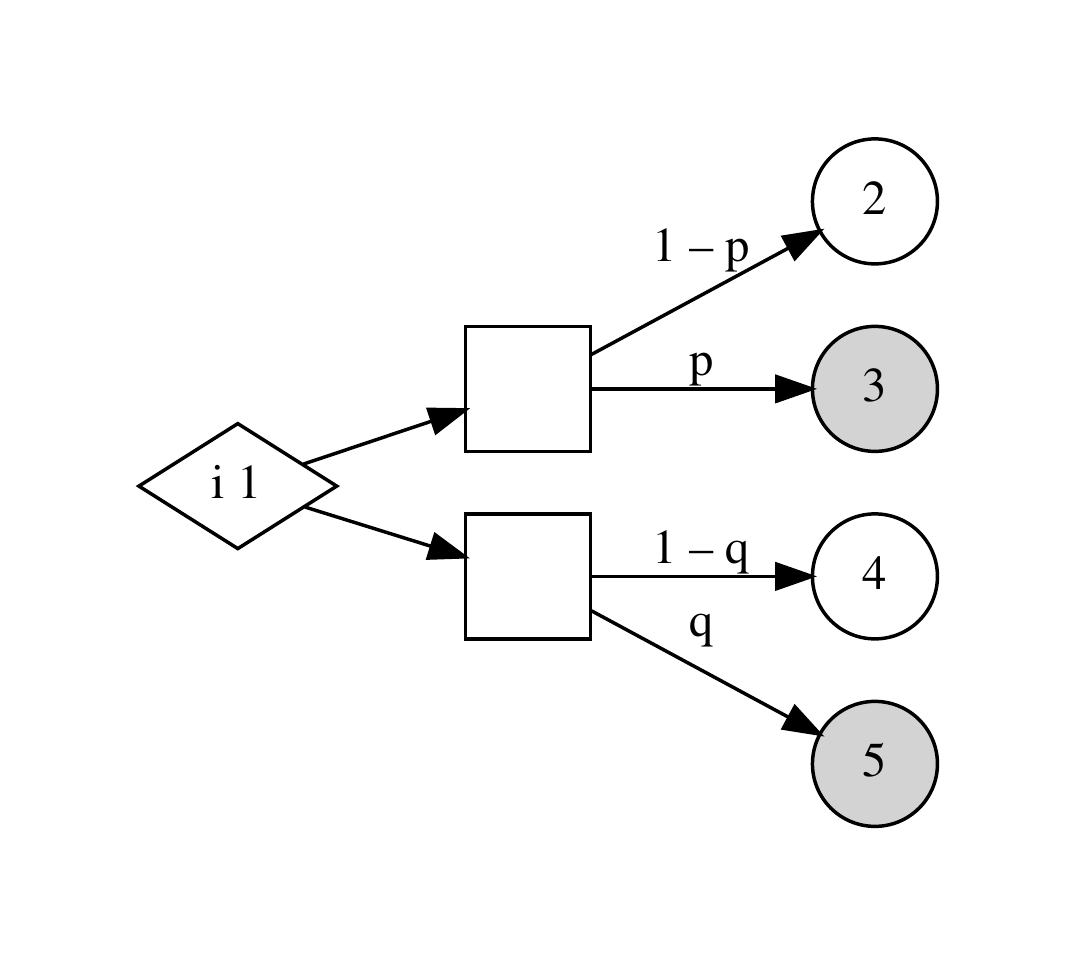}\quad
    {\bf (d)}\includegraphics[width=.45\textwidth]{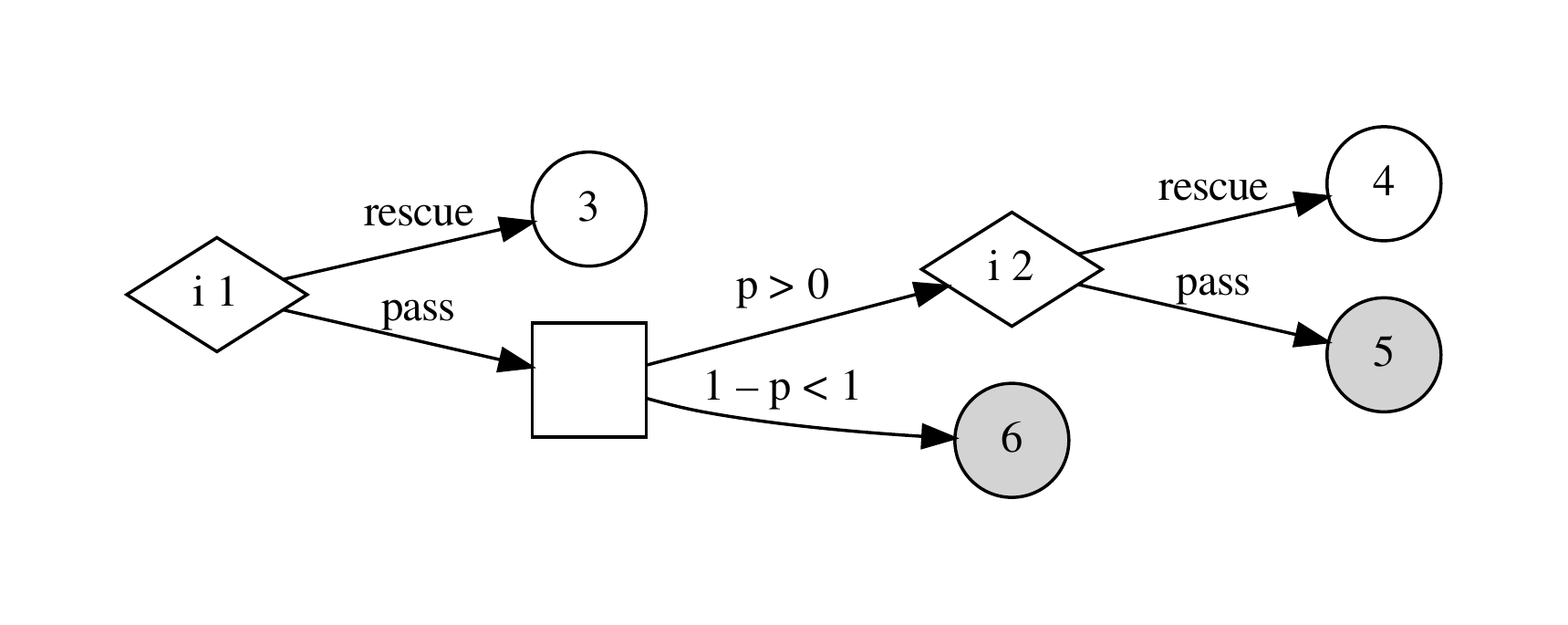}
\caption{\label{fig:paradigmatic}
Multi-agent decision situations that are paradigmatic for the assessment of responsibility,
modelled by a suitable type of decision tree.
Diamonds represent decisions and ambiguities, squares stochastic uncertainty, 
circles outcomes, which are colored grey if ethically undesired.
Dashed lines connect nodes that an agent cannot distinguish when choosing.
(a) Agent $i$ may shoot a prisoner, not knowing whether the gun was loaded (node $v_2$) or not ($v_1$),
leading to the prisoner dead (node $v_6$) or alive ($v_3,v_4,v_5$).
(b) Agents $i,j$ may each throw a stone into a window, not seeing the other's action.
(c) Agent $i$ can choose between two probabilities of an undesired outcome.
(d) Agent $i$ may rescue someone now or, with some probably, later.  
}
\end{figure}

\begin{itemize}
    \item {\bf Load and shoot.} An agent has the choice to shoot at an innocent prisoner or not, not knowing whether the gun was loaded. Represented in Fig.~\ref{fig:paradigmatic}(a).
    \item {\bf Rock throwing.} An agent has the choice to throw a stone into a window or not, not knowing whether another agent already threw a stone before them. Represented in Fig.~\ref{fig:paradigmatic}(b).
    \item{\bf Choosing probabilities.} An agent cannot select an outcome with certainty, but they can influence the probability of a given event. That is, they have the choice between an option where the undesirable outcome has probability $p$ and an option where it has probability $q$. Represented in Fig.~\ref{fig:paradigmatic}(c).
    \item {\bf Hesitation I.} The agent has the choice to either rescue an innocent stranger immediately, or hesitate, in which case they might get another chance at rescuing the innocent stranger at a later stage, but it might also already be too late.\footnote{While this example clearly seems somewhat odd in direct interaction contexts --- imagine a scenario where someone has the choice to save a person from drowning immediately or first finish off their ice-cream knowing that with probability p the other person will hold up long enough so they can still be rescued --- it represents a common issue in climate change mitigation efforts.} Represented in Fig.~\ref{fig:paradigmatic}(d).\\
    If the agent does get a second chance and then decides to rescue the stranger, certain accounts will not assign backwards responsibility to them. However, they did in fact risk the stranger's death, so it can also be argued that they should be held responsible to some degree.
    \item {\bf Hesitation II.} An agent, who is a former lifeguard and thus trained in first aid, passes a stranger who is seemingly having a heart attack. They have the choice to either help immediately by calling an ambulance and keeping up CPR until the ambulance arrives, in which case the stranger survives. Alternatively they can hesitate, but decide again at a later stage whether to help after all. In this case it is not certain whether the stranger will survive. Represented in Fig.~\ref{fig:ind}.\\
    This example is parallel to the one before in the sense that the agent can in a first step hesitate, with an uncertainty determining either before or after their second decision to help after all whether this decision is an option, or whether it is successful. While one might think two consecutive decisions can be considered equivalent to one single combined decision it can be argued in this case that if the agent does not end up helping they failed twice and should thus possibly carry higher responsibility.
    \item {\bf Climate Change.}  
    Humanity (agent $i$) has the choice to either heat up the earth or not, not knowing whether they are in a state of impending heating due to the greenhouse effect or a state of impending cooling due to an onsetting ice age.
    \item {\bf Knowledge gain.} Here Humanity (agent $i$) is again posed before the same issue as in the previous example. But this time they have the added opportunity to learn about which state they are in (impending ice age or not) before deciding on an action.
\end{itemize}

The examples {\em Load and shoot} and {\em Rock throwing} are parallel to one another, both including situations in which the agent might not actually be able to influence the outcome (because either the gun is not loaded so it does not matter whether they shoot or not, or because the other agent already threw a stone that will shatter the window), but they do not know whether they are in this situation or in the one where their action does have an impact. In both cases we argue that the responsibility ascription must take into account the viable option that the agent's action will have/would have had an impact. Therefore, the agent cannot dodge responsibility by referring to this uncertainty. They should be assigned full forward and backward (if they select the possibly harmful action) responsibility. This relates to the discussion about moral luck, and the case for disregarding factors that lie outside of the agent's control is argued in \cite{Nagel1979}: ``Where a significant aspect of what someone does depends on factors beyond his control, yet we continue to treat him in that respect as an object of moral judgment, it can be called moral luck. Such luck can be good or bad. [\dots] If the condition of control is consistently applied, it threatens to erode most of the moral assessments we find it natural to make.''

This also relates to a prominent criticism of the probability raising account for causation, namely that an agent may raise the probability of an event without this event actually occurring as a result, as the probability stayed below 1. Similarly to situations in which the event does not end up occurring due to the actions of others that the agent had no knowledge or influence over, we argue that this should not reduce responsibility ascription but rather be interpreted as a form of `counterfactual' responsibility.

\begin{figure}
    {\bf (a)}\includegraphics[width=.45\textwidth]{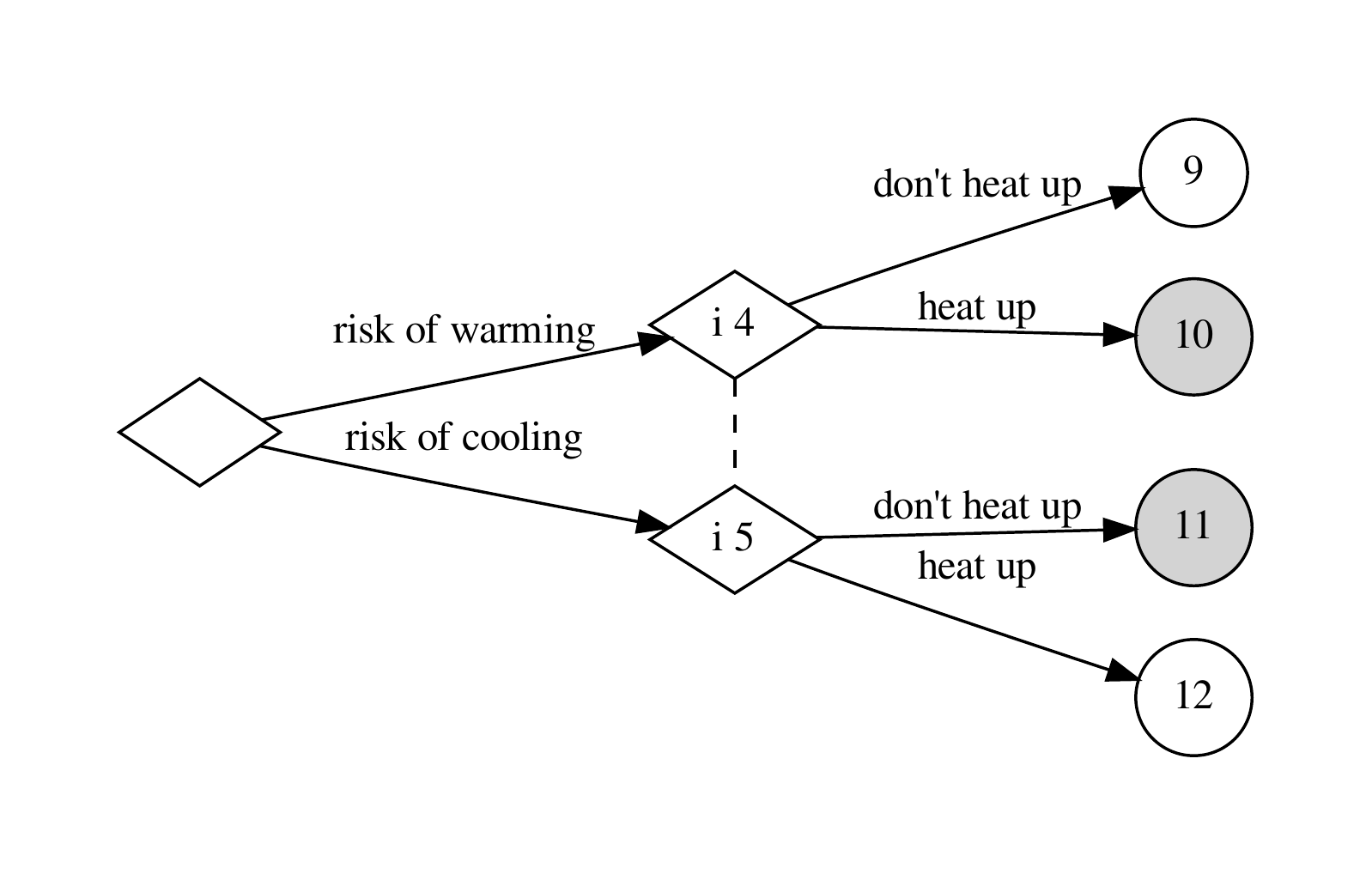}\quad
    {\bf (b)}\includegraphics[width=.5\textwidth]{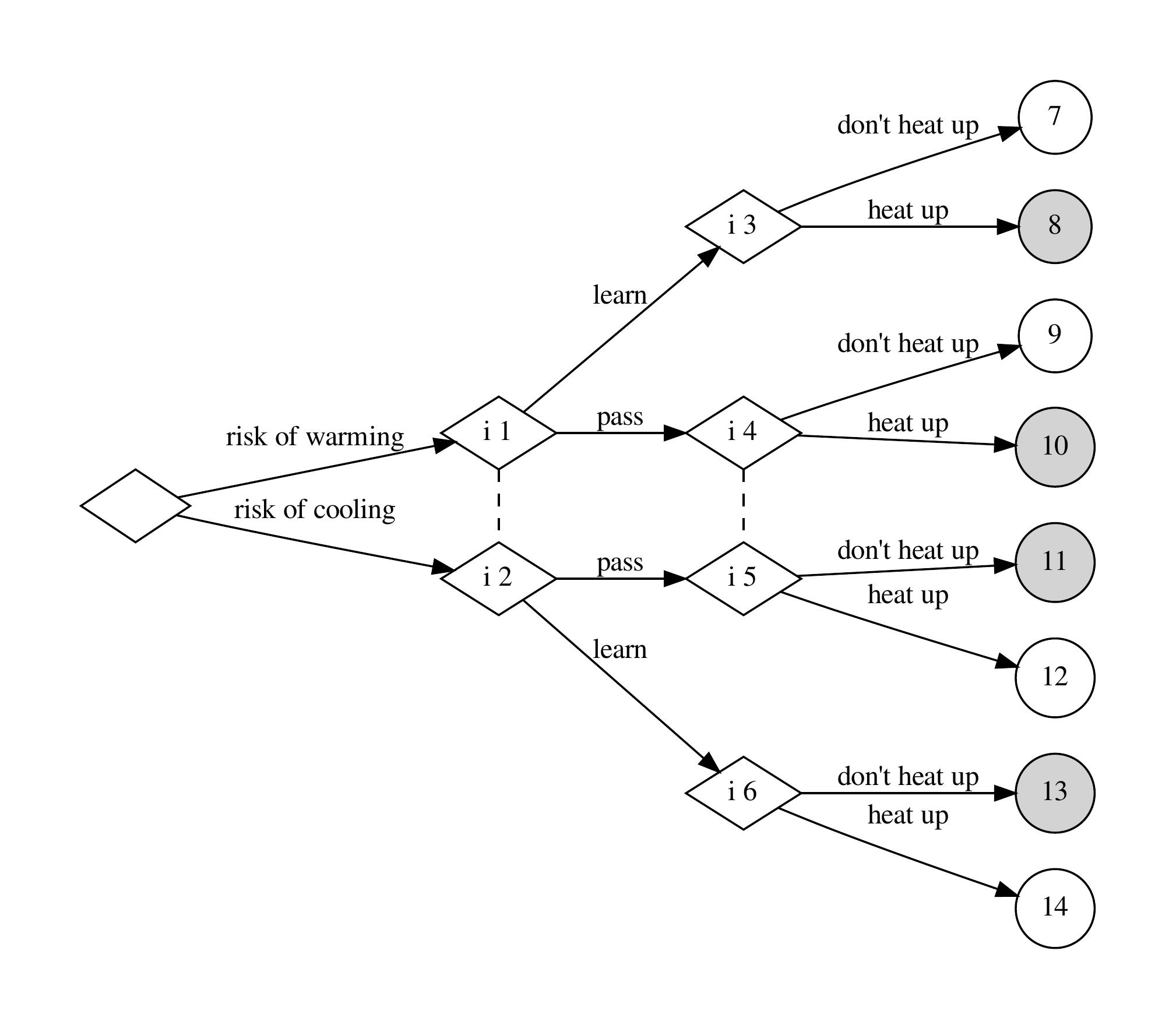}\quad
\caption{\label{fig:learn}
Stylized version of a decision problem related to climate change, 
used to study the effect of options to reduce ambiguity on responsibility.
Humanity (agent $i$) must choose between heating up Earth or not, 
initially not knowing whether there is a risk of global warming or cooling (a),
but potentially being able to acquire this knowledge by learning (b).
While at present, humanity is in node 3, in the 1970's they might rather have been in nodes 1.
}
\end{figure}

 
\paragraph{Structure.} The rest of the paper is structured as follows. We will begin in Sect.~\ref{sec:formalmodel} with a presentation of the proposed framework in which the responsibility functions as well as their desired properties will be formulated. Additionally, we explicate a number of desirable properties that will be important in drawing a difference between the various responsibility functions. In Sect.~\ref{sec:rfunctions} we introduce four different candidate responsibility functions (all differentiated between backward- and forward-looking formulations) and determine which of the axioms they fulfil. Subsequently, in Sect.~\ref{sec:socialchoice} we present a number of voting scenarios known from social choice theory and determine agent's responsibility ascription within these scenarios. In Sect.~\ref{sec:discussion} we discuss selected aspects of our results and finally conclude in Sect.~\ref{sec:conclusion}. 


\section{Formal model}\label{sec:formalmodel}

We start this section by proposing a specific formal framework for the study of responsibility
in multi-agent settings with stochasticity and ambiguities.
It is based on the game-theoretical data structure of a game in extensive form,
which is a multi-agent version of a decision tree,
but with the additional possibility of encoding ambiguity via a special type of node. 
Also, in contrast to games, we do not specify individual payoffs for all outcomes
but only a set of ethically undesired outcomes.
This is sufficient, as we will not apply any game-theoretic analyses referring to rational courses of actions or utility maximisation but rather use this data structure to talk about responsibility assignments.

\subsection{Framework}

We use $\Delta(A)$ to denote the set of all probability distributions on a set $A$,
and use the abbreviations  $A+B\eqdef A\cup B$, $A+a\eqdef A\cup\{a\}$, $A-B\eqdef A\setminus B$, $A-a\eqdef A\setminus\{a\}$.

\paragraph{Trees.}
We define a {\em multi-agent decision-tree with ambiguity} (or shortly, a {\em tree}) to be a structure\\ 
$\T = \langle I,(V_i),V_a,V_p,V_o,E,\sim,(A_v),(c_v),(p_v)\rangle$ 
consisting of:
\begin{itemize}
\item A nonempty finite set $I$ of {\em agents} (or players).
\item For each $i\in I$, a finite set $V_i$ of $i$'s {\em decision nodes}, all disjoint.
      We denote the set of all decision nodes by
      $V_d \eqdef  \bigcup_{i\in I} V_i$.
\item Further disjoint finite sets of nodes:
      a set $V_a$ of {\em ambiguity nodes}, 
      a set $V_p$ of {\em probability nodes},
      and a nonempty set $V_o$ of {\em outcome nodes}.
      We denote the set of all nodes by $V \eqdef  V_d + V_a + V_p + V_o$.
\item A set of directed edges $E\subset V\times V$ so that $(V,E)$ is a directed tree 
      whose leaves are exactly the outcome nodes:
      $$ V_o = \{ v\in V : \not\exists v'\in V ((v,v')\in E) \}. $$
      For all $v\in V-V_o$, let $S_v \eqdef  \{ v'\in V: (v,v')\in E \}$ denote the set of {\em possible successor nodes} of $v$.
\item An {\em information equivalence} relation $\sim$ on $V_d$ so that $v'\sim v\in V_i$ implies $v'\in V_i$.
      We call the equivalence classes of $\sim$ in $V_i$ the {\em information sets} of $i$. 
\item For each agent $i\in I$ and decision node $v\in V_i$, 
      a nonempty finite set $A_v$ of $i$'s possible {\em actions} in $v$,
      so that $A_v = A_{v'}$ whenever $v\sim v'$, 
      and a bijective {\em consequence} function mapping actions to successor nodes, $c_v:A_v\to S_v$.
\item For each probability node $v\in V_p$, a probability distribution $p_v\in\Delta(S_v)$ 
      on the set of possible successor nodes. 
\end{itemize}
Our interpretation of these ingredients is the following:
\begin{itemize}
\item A tree encodes a multi-agent decision situation where certain agents can make certain choices in a certain order,
      and outcome node $v\in V_o$ represents a possible ethically relevant state of affairs 
      that may result from these choices.
\item Each decision node $v\in V_i$ represents a point in time where agent $i$ has the agency to make a decision at free will.
      The elements of $A_v$ are the mutually exclusive choices $i$ can make, including any form of ``doing nothing'',
      and $c_v(a)$ encodes the immediate consequences of choosing $a$ in $v$. 
      Often, $c_v(a)$ will be an ambiguity or probability node to encode uncertain consequences of actions.
\item Probability and ambiguity nodes and information-equivalence are used to represent 
      various types of uncertainty and agents' knowledge at different points in time
      regarding the current state of affairs, immediate consequences of possible actions, 
      future options and their possible consequences, and agents' future knowledge at later nodes. 
      The agents are assumed to always commonly know the tree, and at every point in time to know in which information set they currently are.
      In particular, they know that at any probability node $v\in V_p$,
      the possible successor nodes are given by $S_v$ and have probabilities $p_v(v')$, $v'\in S_v$.
      In contrast, about an ambiguity node $v\in V_a$ they only know that
      the possible successor nodes are given by $S_v$, without being able to rightfully attach probabilities to them. 
      Ambiguity nodes can also be thought of as decision nodes associated to a special agent one might term `nature'.
      
      In contrast to the universal uncertainty at the tree-level encoded by probability and ambiguity nodes, information-equivalence is used to encode uncertainty at the agent level.       
      While in a certain information set of information-equivalent decision nodes, an agent $i$ cannot distinguish between nodes $v\sim v'$ and has the same set of possible actions $A_v = A_v'$.
\item When setting up a tree model to assess some agent $i$'s responsibility, the modeler must carefully decide which actions and ambiguities to include.
      If the modeler follows the basic idea that what matters is what $i$ ``reasonably believes'' in any decision node $v_d$ (as in \cite{baron2016justification}),
      then $A_{v_d}$ should consist of those options that $i$ reasonably believes to have, 
      $S(v)$ for $v\in V_a\cup V_p$ should reflect what possibilities $i$ reasonably beliefs exist at $v$,
      the choice whether $v$ is an ambiguity or probability node should depend on whether $i$ can reasonably believe in certain probabilities of these possibilities,
      and if so, then $p_v$ should reflect those subjective but reasonable probabilities.
      Likewise, if the modeler follows the view that certain forms of ignorance may be a moral excuse (as in \cite{zimmerman2016ignorance}),
      the information equivalence relation $\sim$ should reflect what ignorance of this type the agents have.
\end{itemize}
An ambiguity node whose successors are probability nodes can be used to encode uncertain probabilities
like those reported by the IPCC \cite{ipcc_uncertain} 
or those corresponding to the assumption that ``nature'' uses an Ellsberg strategy \cite{Decerf2019}.

Note that in contrast to some other frameworks, e.g., those using {\em normal}-form (instead of extensive-form) game forms such as \citep{brahamvanhees2018},
our trees do not directly allow for two agents to act at the exact same time point.
Indeed, in a real world in which time is continuous, one action will almost certainly precede another, if only by a minimal time interval. 
Still, as in the theory of extensive-form games, two actions may be considered ``simultaneous'' for the purpose of the analysis 
if they occur so close in time that the later acting player cannot know what the earlier action was,
and this ignorance can easily be encoded by means of information equivalence in a way similar to Fig.\ \ref{fig:majority}.

\paragraph{Events, groups, responsibility functions (RFs).}
As in probability theory, we call each subset $\eps\subseteq V_o$ of outcomes a possible {\em event.}
In the remainder of this paper, we will use $\eps$ to represent an ethically {\em undesirable} event,
such as the death of an innocent person, the occurrence of strong climate change, 
or the election of an extremist candidate, whose probability might be influenced by the agents.

Any nonempty subset $G\subseteq I$ of agents is called a {\em group} in this article.\footnote{Note that we deliberately do not require that a set of agents shares any identity or possesses ways of communication or coordination for an ethical observer to meaningfully attribute responsibility to this ``group''.}

Our main objects of interest are quantitative metrics of degrees of responsibility that we formalise as 
{\em backward-responsibility functions} (BRF) $\R_b$ 
and {\em forward-responsibility functions} (FRF) $\R_f$. 

A BRF maps every combination of tree $\T$, group $G$, event $\eps$, and {\em outcome} node $v\in V_o$ 
to a real number $\R_b(\T,v,G,\eps)$
meant to represent some form of degree of backward-looking (aka ex-post or retrospective) responsibility of $G$ regarding $\eps$ 
in the multi-agent decision situation encoded by $\T$ when outcome $v$ has occurred.

An FRF maps every combination of tree $\T$, group $G$, event $\eps$, and {\em decision} node $v\in V_d$ 
to a real number $\R_f(\T,v,G,\eps)$
meant to represent some form of degree of forward-looking (aka ex-ante) responsibility of $G$ regarding $\eps$ 
in the multi-agent decision situation encoded by $\T$ when in decision node $v$.

If $G=\{i\}$, we also write $\R_{b/f}(\T,v,i,\eps)$.
Whenever any of the arguments $\T$, $v$, $G$, $\eps$ are kept fixed and are thus obvious from the context,
we omit to explicate them when writing $\R_{b/f}$ or any of the auxiliary functions defined below.

\paragraph{Graphical representation.}
As exemplified in Fig.\ \ref{fig:paradigmatic}, we can represent a tree $\T$ and event $\eps$ graphically as follows.
Edges are arrows, 
decision nodes are diamonds labelled by agents, with arrows labelled by actions, 
ambiguity nodes are unlabelled diamonds,
probability nodes are squares with arrows labelled by probabilities,   
and outcome nodes are circles, filled in grey if the outcome belongs to $\eps$.
Finally, information equivalence is indicated by dashed lines connecting or surrounding the equivalent nodes. 

\paragraph{Auxiliary notation.}
The set of decision nodes of a group $G\subseteq I$ is $V_G \eqdef  \bigcup_{i\in G}V_i$.
To ease the definition of ``scenario'' below we denote the set of non-probabilistic uncertainty nodes other than $V_G$ (i.e., non-$G$ decision and ambiguity nodes) by $V_{-G} \eqdef  V_d - V_G + V_a$.\footnote{This can be thought of as nodes where someone who is not part of group G --- another agent or Nature --- takes a decision.}

If $v'\in S_v$, we call $P(v') \eqdef v$ the {\em predecessor} of $v'$.
Let $v_0 \in V$ be the {\em root} node of $(V,E)$, i.e., the only node without predecessor.
The {\em history} of $v\in V$ is then $H(v) \eqdef  \{ v, P(v), P(P((v)), \dots, r \}$, where $r$ is the root node of $(V,E)$.
In the other direction, we call $B(v) \eqdef  \{v'\in V: v\in H(v')\}$ the {\em (forward) branch} of $v$.
Taking into account information equivalence, we also define the {\em information branch} of $v$ as
$B^\sim(v) \eqdef  \bigcup_{v'\sim v} B(v')$.

If $v\in V$, $v_d\in H(v)\cap V_d$, and $c_{v_d}(a)\in H(v)$, we call $C_{v_d}(v) \eqdef  a$ the {\em choice} at $v_d$ that ultimately led to node $v$.

A node $v\in V_d$ with $\{v':v'\sim v\}=\{v\}$ is called a {\em complete information} node.

\paragraph{Strategies, scenarios, likelihoods.}

We call a function $\sigma:V_G^\sigma\to \bigcup_{v_d\in V_G^\sigma}A_{v_d}$ 
that chooses actions $\sigma(v_d)\in A_{v_d}$ for some set $V_G^\sigma$ of $G$'s decision nodes a {\em partial strategy} for $G$ at $v$ iff 
$v\in V$,  
$V_G^\sigma\subseteq V_G\cap B^\sim(v)$,
$\sigma(v_d)=\sigma(v_d')$ whenever $v_d\sim v_d'$,
and
$V_G^\sigma\cap B^\sim(c_{v_d}(a))=\emptyset$
for all $v_d\in V_G^\sigma$ and $a\in A_{v_d}-\sigma(v_d)$.
The latter condition says that $\sigma$ does not specify actions for decision nodes that become unreachable by earlier choices made by $\sigma$.
A {\em strategy} for $G$ at $v$ is a partial strategy with a maximal domain $V_G^\sigma$.
This means that a strategy specifies actions for all decision nodes that can be reached from the information set containing $v$ given the strategy.

Let $\Sigma(\T,v,G)$ (or shortly $\Sigma(v)$ if $\T,G$ are fixed) be the set of all those strategies.
For $\sigma\in\Sigma(\T,v,G)$, let 
$$ V_o^\sigma \eqdef \{ v_o\in B^\sim(v)\cap V_o : C_{v_d}(v_o) = \sigma(v_d)\text{~for all~}v_d\in B^\sim(v)\cap H(v_o)\cap V_G \}, $$
i.e., the set of possible outcomes when $G$ follows $\sigma$ from $v$ on.

Complementary, consider a function $\zeta:V^\zeta\to\bigcup_{v'\in V^\zeta}S_{v'}$
that chooses successor nodes $\zeta(v')\in S_{v'}$ for a set $V^\zeta$ of 
ambiguity or others' decision nodes, and some node $v_\zeta\in V_d$.
Then we call $\zeta$ a {\em partial scenario} for $G$ at $v$ iff
$v\in V$,
$v_\zeta=v$ or $v_\zeta\sim v$,
$V^\zeta\subseteq V_{-G}\cap B(v^\zeta)$, 
$\zeta(v')=c_{v'}(a)$ and $\zeta(v'')=c_{v''}(a)$ for some $a\in A_{v'}$ whenever $v'\sim v''\in V^\zeta$,
and $V^\zeta\cap B^\sim(v'')=\emptyset$ for all $v'\in V^\zeta$ and $v''\in S(v')-\zeta(v')$.
The latter condition says that $\zeta$ does not specify successors for nodes becoming unreachable under $\zeta$.
A {\em scenario} for $G$ at $v$ is a partial scenario with a maximal domain $V^\zeta$.
This means that a scenario specifies successors for all ambiguity and others' decision nodes that can be reached from $v$ or the information set containing $v$ given the scenario.

Let $Z^\sim(\T,v,G)$ (or shortly $Z^\sim(v)$) be the set of all scenarios at $v$
and $Z(\T,v,G)\subseteq Z^\sim(\T,v,G)$ (or shortly $Z(v)$) that of all scenarios at $v$ with $v_\zeta=v$.

Each strategy-scenario pair $(\sigma,\zeta)\in \Sigma(v)\times Z^\sim(v)$ 
induces a Markov process on $B^\sim(v)$ leading to a {\em prospect}, 
i.e., a probability distribution $\pi_{v,\sigma,\zeta}\in\Delta(V_o\cap B^\sim(v))$ on the potential future outcome nodes, 
that can be computed recursively in the following straightforward way:
\begin{align}
    \psi(v_\zeta) &= 1, \\
    \psi(v'') &= \psi(v_d)\quad \text{if~} [v_d\in V_G\wedge v''=c_{v_d}(\sigma(v_d))]\vee[v'\in V_{-G}\wedge v''=\zeta(v')], \\
    \psi(v'') &= \psi(v') p_{v'}(v'')\quad \text{for~} v'\in V_p, v''\in S_{v'}, \\
    \psi(v'') &= 0\quad \text{for all other~} v''\in B^\sim(v), \\
    \pi_{v,\sigma,\zeta}(v_o) &= \psi(v_o) \quad \text{for all~} v_o\in V_o\cap B^\sim(v).
\end{align}
Let us denote the resulting {\em likelihood} of $\eps$ by 
$$ \ell(\eps|v,\sigma,\zeta) \eqdef \sum_{v_o\in\eps}\pi_{v,\sigma,\zeta}(v_o). $$

\subsection{Axioms}
Following an axiomatic approach similar to what social choice theory does for group decision methods and welfare functions,
we study RFs by means of a number of potentially desirable properties formalized as {\em axioms.}

In the main text, we focus on a selection of axioms 
which turn out to motivate or distinguish between certain variants of RFs 
that we will develop in the next section and then apply to social choice mechanisms.
In the Appendix, a larger list of plausible axioms is assembled and discussed.

All studied RFs fulfill a number of basic symmetry axioms such as anonymity (treating all agents the same way),
and a number of independence axioms such as the independence of branches with zero probability,
and, more notably, also the following two axioms:
\begin{description}
\item[(IOA)] {\em Independence of Others' Agency.}
  If $i\in I - G$, and some of $i$'s decision nodes $v_d\in V_i$ is turned into an ambiguity node $v_a$ with $S_{v_a}=S_{v_d}$,
  then $\R(G)$ remains unchanged 
  (i.e., it is irrelevant whether uncertain consequences are due to choices of other agents 
  or some non-agent mechanism with ambiguous consequences).
\item[(IGC)] {\em Independence of Group Composition.}
  If $i,i'\in G$ and all occurrences of $i'$ are replaced in $\T$ by $i$, $\R(G)$ remains unchanged.
\end{description}
Note that these two conditions preclude dividing a group's responsibility equally between its members or following other agent- or group-counting approaches similar to Banzhaf's or other power indices.

\begin{figure}
    \centering\includegraphics[width=.55\textwidth]{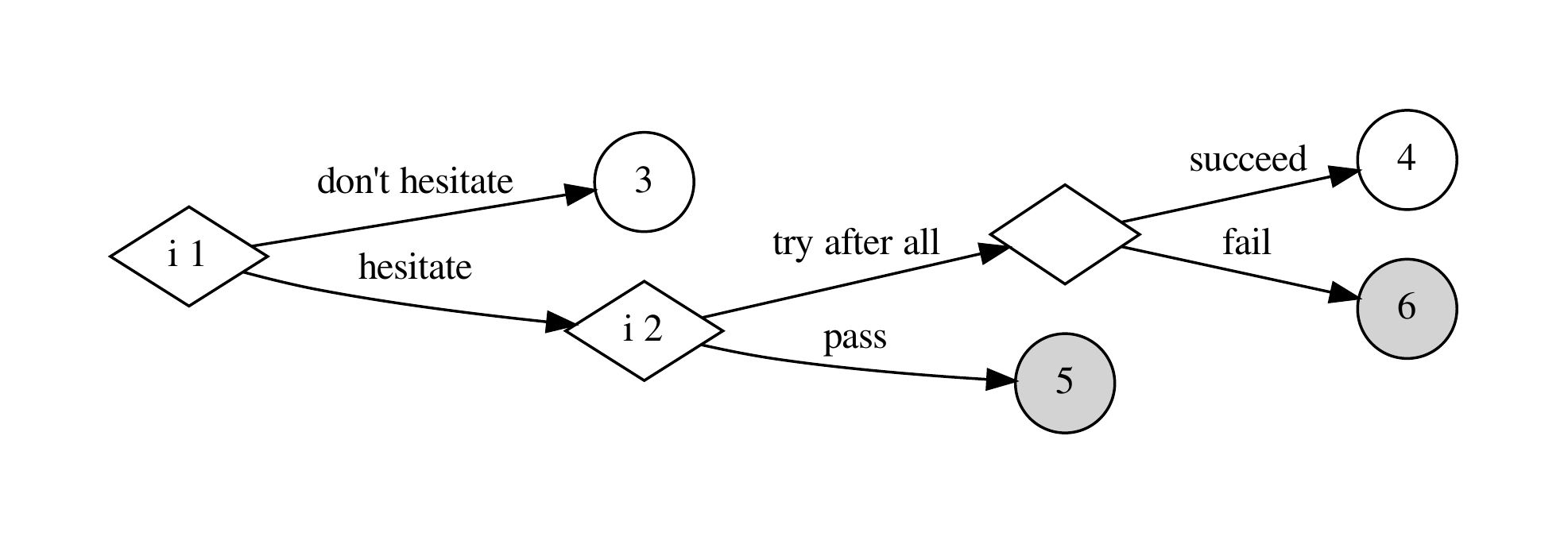}
\caption{\label{fig:ind}
Situation related to the Independence of Nested Decisions (IND) axiom. The agent sees someone having a heart attack and may either try to rescue them without hesitation, applying CPU until the ambulance arrives, or hesitate and then reconsider and try rescuing them after all, in which case it is ambiguous whether the attempt can still succeed.}
\end{figure}

\begin{figure}
    \centering\includegraphics[width=.7\textwidth]{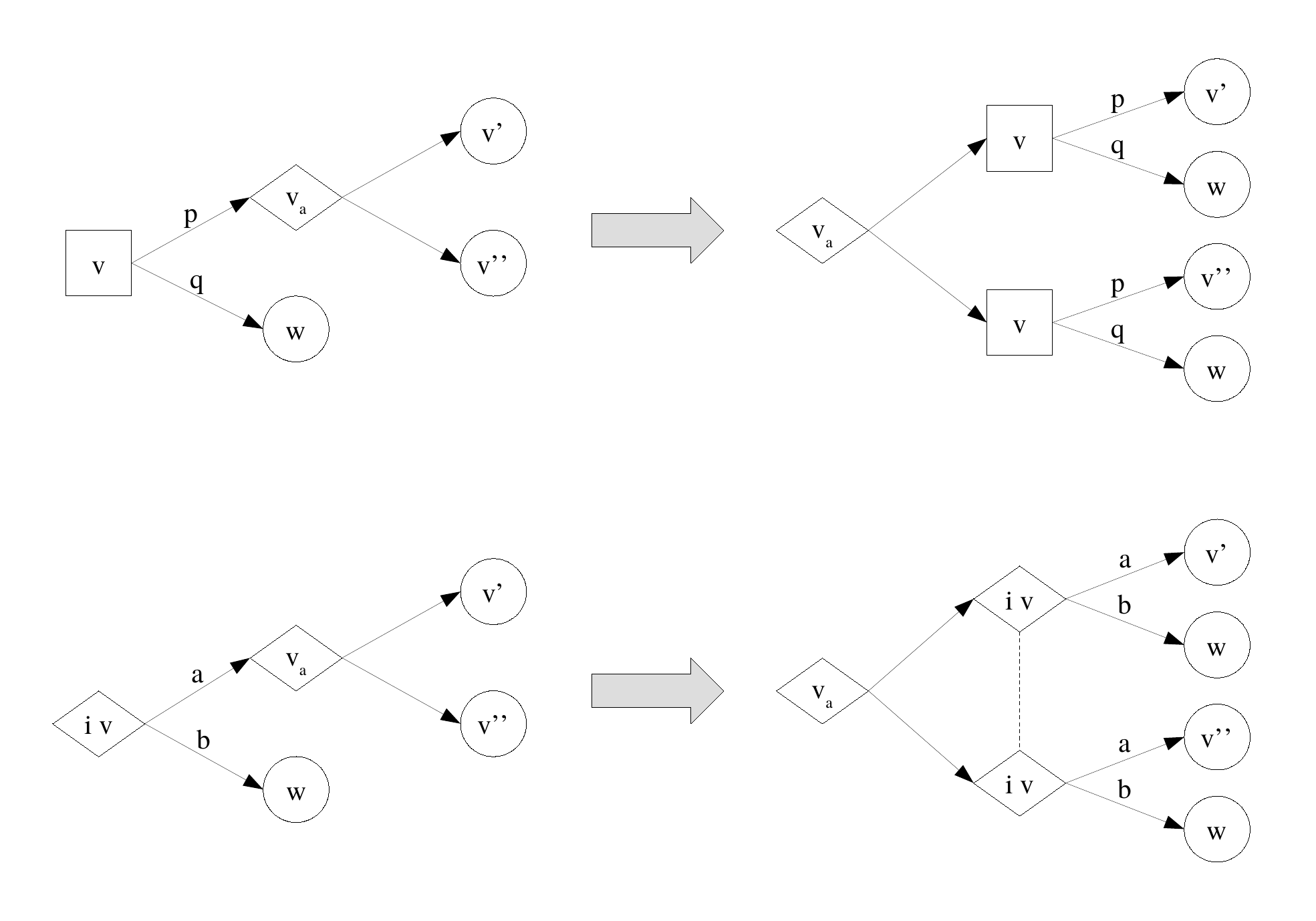}
\caption{\label{fig:iat}
Explanation of the ``pulling back'' transformation described in the (IAT) axiom. 
Top: pulling back an ambiguity node before a probability node;
bottom: pulling back an ambiguity node before a decision node, leading to information equivalence.}
\end{figure}

The first two axioms that only some of our candidate RFs will fulfill are the following:  
\begin{description}
\item[(IND)] {\em Independence of Nested Decisions.}
  If a complete-information decision node $v_d\in V_i$ is succeeded via some action $a\in A_{v_d}$ 
  by another complete-information decision node $v_d'=c_{v_d}(a)\in V_i$ of the same agent, 
  then the two decisions may be treated as part of a single decision, 
  i.e., $v_d'$ may be pulled back into $v_d$: $v_d'$ may be eliminated, $S_{v_d'}$ added to $S_{v_d}$, 
  $\{a\}\times A_{v_d'}$ added to $A_{v_d}$, and $c_{v_d}$ extended by $c_{v_d}(a,a') = c_{v'_d}(a')$ for all $a'\in A_{v_d'}$.
\item[(IAT)] {\em Independence of Ambiguity Timing.}
  Assume some probability node $v\in V_p$ or complete-information decision node $v\in V_d$ 
  is succeeded by an ambiguity node $v_a\in V_a\cap S_{v}$.
  Let $B(v)$, $B(v_a), B(v')$ be the original branches of the tree $(V,E)$ starting at $v$, $v_a$ and any $v'\in S_{v_a}$.
  For each $v'\in S_{v_a}$, 
  let $B'(v')$ be a new copy of the original $B(v)$ in which the subbranch $B(v_a)$ is replaced by a copy of $B(v')$;
  let $f(v')$ be that copy of $v$ that serves as the root of this new branch $B'(v')$.
  If $v\in V_d$, put $f(v')\sim f(v'')$ for all $v',v''\in S_{v_a}$
  Let $B'(v_a)$ be a new branch starting with $v_a$ and then splitting into all these new branches $B'(v')$.
  Then $v_a$ may be ``pulled before'' $v$ by replacing the original $B(v)$ by the new $B'(v_a)$,
  as exemplified in Fig.\ \ref{fig:iat}.
\end{description}
(IND) may seem plausible if one imagines, say, a decision to turn either left or right directly followed by a decision to stop at 45 or 90 degrees rotation,
since these two may more naturally be considered a single decision between four possible actions, turning 90 or 45 degrees left or right.
But in the situation of Fig.\ \ref{fig:ind}, it may rather seem that when hesitating and then passing, 
$i$ has failed twice in a row, which should perhaps be assessed differently from having failed only once.


When using an RF with all the above properties to assess responsibility of a particular group of agents, 
one can ``reduce'' the original tree to one that has
only a single agent (representing the whole group, all other actions being represented as simple ambiguity).
If one accepts a number of similar further axioms listed in the Appendix, 
one can also assume the reduced tree
has only properly branching non-outcome nodes,
has at most one ambiguity node and only as its root node, 
and has no two consecutive probability nodes and no zero probability edges.

The next pair of axiom state that responsibility must react in the right direction under certain modifications:
\begin{description}
\item[(GSM)] {\em Group Size Monotonicity.}
  If $G\subseteq G'$ then $\R(G)\le \R(G')$
  (i.e., larger groups have no less responsibility).
\item[(AMF)] {\em Ambiguity Monotonicity of Forward Responsibility.}
  If, from an ambiguity node $v_a\in V_a$, we remove a possibility $v'\in S_{v_a}$ and its branch $B(v')$, then forward responsibility $\R_f(v)$ in any remaining node $v\notin B(v')$ does not increase.
\end{description}
In other words, (AMF) requires that increasing ambiguity should not lower forward responsibility (because that might create an incentive to not reduce ambiguity).

The next three axioms set lower and upper bounds for responsibility,
the first taking up a condition from \citep{brahamvanhees2018}:
\begin{description}
\item[(NRV)] {\em No Responsibility Voids.}
  If there is no uncertainty, $V_a=V_p=\emptyset$,
  and if $\eps\neq V_o$,
  then for each undesired outcome $v_o\in\eps$, 
  some group $G\subseteq I$ is at least partially responsible, $\R_b(v_o,G) > 0$.
\item[(NUR)] {\em No Unavoidable Backward Responsibility.} 
  Each group $G\subseteq I$ must have an original strategy $\sigma\in\Sigma(v_0,G)$ 
  that is guaranteed to avoid any backward responsibility,
  i.e., so that $\R_b(v_o,G) = 0$ for all $v_o\in V_o^\sigma$.
\item[(MBF)] {\em Maximal Backward Responsibility Bounds Forward Responsibility.} 
  For all $v,G$, there must be $\sigma\in\Sigma(v,G)$ and $v_o\in V_o^\sigma$ 
  so that $R_b(v_o,G) \ge \R_f(v,G)$
  (i.e., forward responsibility is bounded by potential backward responsibility).
\end{description}

Finally, we consider four axioms that require certain assessments in the paradigmatic situations from Fig.\ \ref{fig:paradigmatic} which are closely related to questions of moral luck \citep{Nagel1979,Andre1983,Tong2004},
reasonable beliefs \cite{baron2016justification}, and ignorance as an excuse \cite{zimmerman2016ignorance}:
\begin{description}
\item[(NFT)] {\em No Fearful Thinking.}
  With $\T$ and $\eps$ as depicted in Fig.\ \ref{fig:paradigmatic}(b), 
  $\R_f(v_1,\{i\})=1$ since $i$'s action makes a difference even though she thinks acting might not help,
  since it would be unreasonable to believe this must be the case.
\item[(NUD)] {\em No Unfounded Distrust.}
  With $\T$ and $\eps$ as depicted in Fig.\ \ref{fig:paradigmatic}(b), 
  $\R_f(v_2,\{i\})=1$ since $i$ cannot know that acting cannot help.
\item[(MFR)] {\em Multicausal Factual Responsibility.} 
  With $\T$ and $\eps$ as depicted in Fig.\ \ref{fig:paradigmatic}(b), 
  $\R_b(v_6,\{i\})=1$ since $i$'s action was necessary even though not sufficient.
\item[(CFR)] {\em Counterfactual Responsibility.} 
  With $\T$ and $\eps$ as depicted in Fig.\ \ref{fig:paradigmatic}(a), 
  $\R_b(v_4,\{i\})=1$ since $i$ could not know that her action would not cause $\eps$ so she must reasonably have taken into account that it might.
\end{description} 

Before turning to the definition of candidate RFs and study their axiom compliance, we briefly mention that while there obviously exist certain logical relationships between subsets of the above axioms (and the further axioms listed in the Appendix), they are not the scope of this article.

\section{Candidate responsibility functions}\label{sec:rfunctions}

Here we will introduce four pairs of responsibility functions $(\R_f, \R_b)$ 
that fulfill most of the above axioms but each also violate a few,
and a reference function $\R_b^0$ related to strict causation.

These candidate responsibility functions will measure degrees of responsibility in terms of 
differences in likelihoods between available strategies in all possible scenarios.

To define them, we need some additional auxiliary notation and terminology.
For now, let us keep $\T$, $G$, and $\eps$ fixed and drop them from notation.

Since the below definitions typically involve several nodes, we denote the decision node at which $\R_f$ is evaluated by $v_d\in V_d$, the outcome node at which $\R_b$ is evaluated by $v_o\in V_o$, and other nodes by $v,v'\in V$ so that $v$ comes before $v'$ (i.e., $v\in H(v')$, $v'\in B(v)$).

\paragraph{Benchmark variant: strict causation.}

The most straightforward definition of a backward responsibility function in our framework that resembles the strict causation view, as employed for example in the most basic way of `seeing to it that', is to set
$\R_b^0(v_o)\eqdef 1$ iff there is a past node $v\in H(v_o)$ at which $\eps$ was certain, $B(v)\cap V_o\subseteq\eps$,
directly following a decision node $v_d=P(v)\in V_G$ at which $\eps$ was not certain, $B(v_d)\cap V_o\not\subseteq\eps$,
and to put $\R_b^0(v_o)\eqdef 0$ otherwise.

It is easy to see that given $v_o$, there is at most one such $v_d$ regardless of $G$, 
and exactly those $G$ are deemed responsible which contain the agent choosing at $v_d$, i.e., for which $v_d\in V_G$.



\subsection{Variant 1: measuring responsibility in terms of causation of increased likelihood}

The rationale for this variant, which tries to translate the basic idea of the stit approach into a probabilistic context,
is that backward responsibility can be seen as arising from 
having caused an increase in the guaranteed likelihood of an undesired outcome.

\paragraph{Guaranteed likelihood, caused increase, backwards responsibility.}
We measure the {\em guaranteed likelihood} of $\eps$ at some node $v\in V$ by the quantity
\begin{align}\label{eq:gamma}
   \gamma(v) &\eqdef  \min_{\sigma\in\Sigma(v)} \min_{\zeta\in Z(v)} \ell(\eps|v,\sigma,\zeta). 
\end{align}

We measure the {\em caused increase in guaranteed likelihood} in choosing $a\in A_{v_d}$ at decision node $v_d\in V_d$ by the difference
\begin{align}\label{eq:Dgamma}
    \Delta\gamma(v_d,a) &\eqdef \gamma(c_{v_d}(a)) - \gamma(v_d).
\end{align}
Note that since $v_d\in V_d$ rather than $v_d\in V_p$, we have $\Delta\gamma(v_d)\ge 0$.

To measure $G$'s {\em backward responsibility} regarding $\eps$ in outcome node $v_o\in V_o$,
in this variant we take their aggregate caused increases over all choices $C_{v_d}(v_o)$ taken by $G$ that led to $v_o$,
\begin{align}
    \R_b^1(v_o) &\eqdef \sum_{v_d\in H(v_o)\cap V_G} \Delta\gamma(v_d,C_{v_d}(v_o)).
\end{align}

\paragraph{Maximum caused increase, forward responsibility.}
Finally, to measure $G$'s {\em forward responsibility} regarding $\eps$ in decision node $v_d\in V_G$,
we take the maximal possible caused increase,
\begin{align}
    \R_f^1(v_d) &\eqdef \max_{a\in A_{v_d}} \Delta\gamma(v_d,a).
\end{align}


At this point, we notice to potential drawbacks of this variant.
For one thing, it fails (IAT), mainly because it does not take into account any information equivalence and thus depends too much on subtle timing issues that the agents information does not depend on and that hence any responsibility assessments should maybe also not depend on.
On the other hand, it is in a sense too ``optimistic'' by allowing agents to ignore the possibility that their action {\em might} make a negative difference if this is not guaranteed to be the case.
The next variant tries to resolve these two issues.

\subsection{Variant 2: measuring responsibility in terms of increases in minimax likelihood}

\begin{figure}
\centering
\includegraphics[width=.35\textwidth]{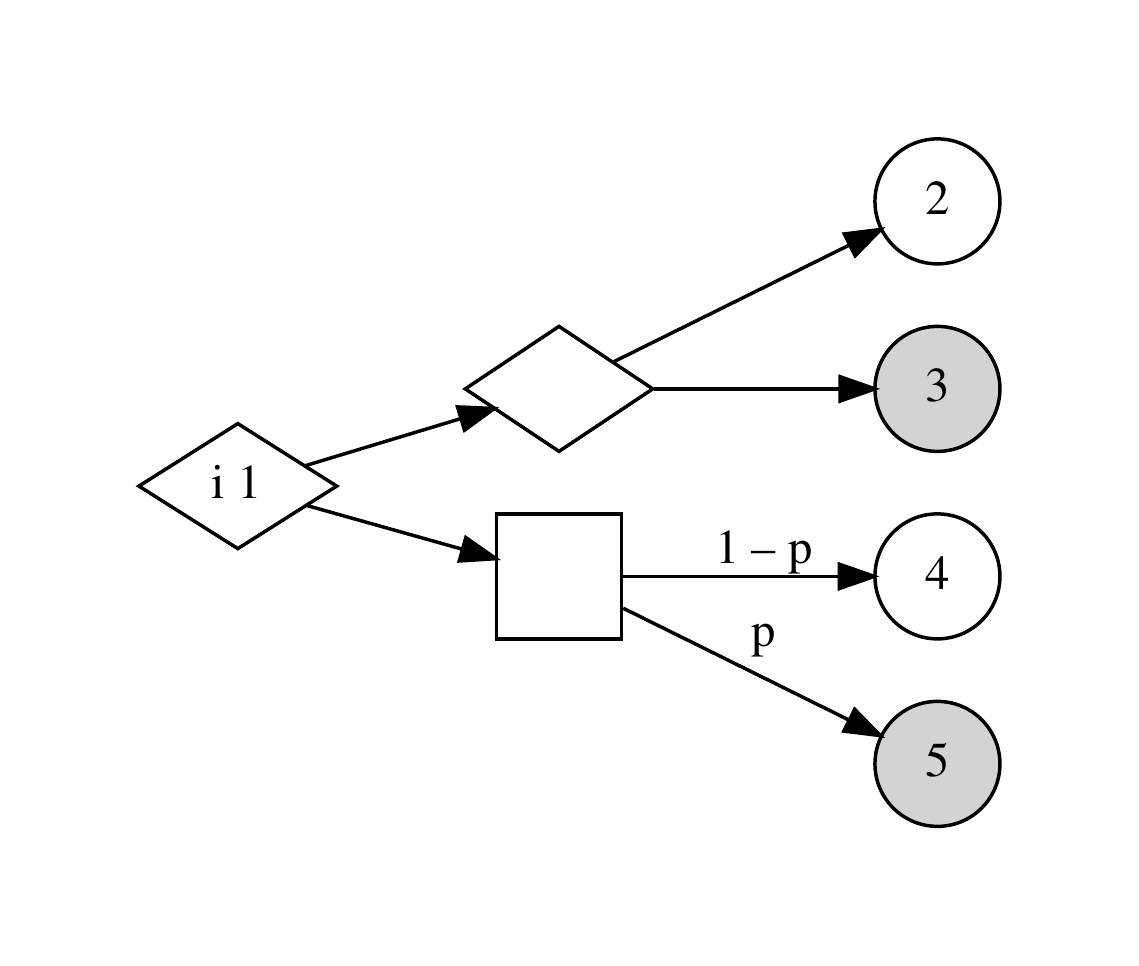}
\caption{\label{fig:ambig}
Situation related to ambiguity aversion in which the complementarity of variants 1 and 2 of our responsibility functions can be seen.
The agent must choose between an ambiguous course and a risky course. The ambiguous course seems the right choice in variant 1 since it does not increase the guaranteed likelihood of a bad outcome, which remains zero, while the risky course seems right in variant 2 since it reduces the minimax likelihood of a bad outcome from 1 to $p$.}
\end{figure}

This variant is in a sense the opposite of variant 1 with respect to its ambiguity attitude. 
To understand their relationship, consider the tree in Fig.\ \ref{fig:ambig} which shows that variant 1 can be interpreted as suggesting an ambiguity-affine strategy while variant 2 suggests an ambiguity-averse strategy.

In this variant, the rationale is that backward responsibility can be seen as arising from having deviated
from behaviour that would have seemed optimal in minimizing the {\em worst-case} (rather than the guaranteed)
likelihood of an undesired outcome in view of the information available at the time of the decision.
In defining the worst-case, however, we assume a group $G$ can plan and commit to optimal future behaviour, 
so some of the involved quantities are in now terms of strategies $\sigma$ rather than actions $a$.

\paragraph{Worst case and minimax likelihoods.}
$G$'s {\em worst-case likelihood} of $\eps$ at any node $v\in V$ given some strategy $\sigma\in\Sigma(v)$ is given by
\begin{align}
    \lambda(v,\sigma) &\eqdef \max_{\zeta\in Z^\sim(v)} \ell(\eps|v,\sigma,\zeta).
\end{align}
$G$'s {\em minimax likelihood} regarding $\eps$ at $v$ is the smallest achievable worst-case likelihood,
\begin{align}\label{eq:mu}
    \mu(v) &\eqdef \min_{\sigma\in\Sigma(v)} \lambda(v,\sigma)
    = \min_{\sigma\in\Sigma(v)} \max_{\zeta\in Z^\sim(v)} \ell(\eps|v,\sigma,\zeta).
\end{align}
Note that (\ref{eq:mu}) differs from (\ref{eq:gamma}) not only in using a maximum but also in taking into account possible ignorance about the true node by using $Z^\sim$ instead of $Z$.

\paragraph{Caused increase, backward responsibility.}
We measure $G$'s {\em caused increase in minimax likelihood} 
in choosing $a\in A_{v_d}$ at node $v_d\in V_G$ by taking the difference
\begin{align}\label{eq:Dmu}
    \Delta\mu(v_d,a) &\eqdef \jobst{\max_{v_d'\sim v_d}}\mu(c_{\jobst{v_d'}}(a)) - \mu(v_d) \ge 0,
\end{align}
\jobst{again now taking information equivalence into account.}
Similar to before, to measure $G$'s {\em backward responsibility} regarding $\eps$ in node $v_o$,
we here take their aggregate caused increases in minimax likelihood,
\begin{align}
    \R_b^2(v_o) &\eqdef \sum_{v_d\in H(v_o)\cap V_G} \Delta\mu(v_d,C_{v_d}(v_o)).
\end{align}

\paragraph{Maximum caused increase, forward responsibility.}
In analogy to variant 1, to measure $G$'s {\em forward responsibility} regarding $\eps$ in node $v_d\in V_G$,
we take the maximal possible caused increase in minimax likelihood, 
\begin{align}
    \R_f^2(v_d) &\eqdef \max_{a\in A_{v_d}} \Delta\mu(v_d,a).
\end{align}

While this variant seems well related to the maximin-type of analysis known from early game theory, it still fails (NUD) and (MFR), both because it is now in a sense too ``pessimistic'' by allowing agents to ignore the possibility that their action might make a positive difference.

%
%
%

\subsection{Variant 3: measuring responsibility in terms of influence and risk-taking}

While variants 1 and 2 can be interpreted as measuring the deviation from a single optimal strategy
that minimizes either the guaranteed (best-case) or the worst-case likelihood of a bad outcome taking into account all ambiguities,
our next variant is based on families of scenario-dependent optimal strategies.
In this way, it partially manages to avoid being too optimistic or too pessimistic
and thereby fulfil both (MFR) like variant 1 and (CFR) like variant 2. 
The main idea is that backward responsibility arises from taking risks to not avoid an undesirable outcome.

\paragraph{Optimum, shortfall, risk, backward responsibility.}
Given a scenario $\zeta\in Z^\sim(v)$ at any node $v\in V$, 
the {\em optimum} $G$ could achieve for avoiding $\eps$ at that node in that scenario is the minimum likelihood over $G$'s strategies at $v$,
\begin{align}
    \omega(v,\zeta) &\eqdef  \min_{\sigma\in\Sigma(v)} \ell(\eps|v,\sigma,\zeta).
\end{align}
So let us measure $G$'s hypothetical {\em shortfall} in avoiding $\eps$ in scenario $\zeta$ 
due to their choice $a\in A_{v_d}$ at node $v_d\in V_G$
by the difference in optima
\begin{align}
    \Delta\omega(v_d,\zeta,a) &\eqdef  
    \omega(c_{\jobst{v_\zeta}}(a),\zeta) - \omega(v_d,\zeta) \ge 0.
\end{align}
Then then {\em risk taken} by $G$ in choosing $a$ is the maximum shortfall over all scenarios at $v_d$,
\begin{align}
    \rho(v_d,a) &\eqdef  \max_{\zeta\in Z^\sim(v_d)} \Delta\omega(v_d,\zeta,a).
\end{align}
To measure $G$'s {\em backward responsibility} regarding $\eps$ in node $v_o\in V_o$,
we now take their aggregate risk taken over all choices they made,
\begin{align}
    \R_b^3(v_o) &\eqdef  \sum_{v_d\in H(v_o)\cap V_G} \rho(v_d,C_{v_d}(v_o)).
\end{align}

\paragraph{Influence, forward responsibility.}
Regarding forward responsibility, we test a different approach than before,
which is simpler but less strongly linked to backward responsibility.
The rationale is that since $G$ does not know which scenario applies, 
they must take into account that their actual influence on the likelihood of $\eps$
might be as large as the maximum of this over all possible scenarios, 
so the larger this value is the more careful $G$ need to make their choices.

Let us measure $G$'s {\em influence} regarding $\eps$ in scenario $\zeta$ at any node $v\in V$ 
by the range of likelihoods spanned by $G$'s strategies at $v$,
\begin{align}
    \Delta\ell(v,\zeta) &\eqdef  \max(L) - \min(L), \\
    L &\eqdef  \{ \ell(\eps|v,\sigma,\zeta) : \sigma\in\Sigma(v) \}.
\end{align}
To measure $G$'s {\em forward responsibility} regarding $\eps$ at node $v_d\in V_G$,
we this time simply take their maximum influence over all scenarios at $v_d$,
\begin{align}
    \R_f^3(v_d) &\eqdef \max_{\zeta\in Z^\sim(v_d)} \Delta\ell(v_d,\zeta).
\end{align}

A main problem with $\R_b^3$ is that it fails (NUR), so that in situations like Fig.\ \ref{fig:learn}(a), 
it will assign full backward responsibility no matter what $i$ did.
This ``tragic'' assessment arises because in such situations, there is no weakly dominant strategy that is optimal in all scenarios,
hence risk-taking cannot be avoided.

\subsection{Variant 4: measuring responsibility in terms of negligence}

In our final variant, we turn the ``tragic'' assessments of variant 3 into ``realistic'' ones, 
making it fulfil (NUR), 
by using risk-minimizing actions as a reference,
but at the cost of losing compliance with (NRV).
We also return to the original idea of basing forward responsibility on potential backward responsibility applied in variants 1 and 2 to fulfil (MBF),
but at the cost of losing compliance with (AMF).

\paragraph{Risk-minimizing action, negligence, backward responsibility.}
The {\em minimal risk} and set of {\em risk-minimizing actions} of $G$ in decision node $v_d\in V_G$ is
\begin{align}\label{eq:alpha}
    \underline\rho(v_d) &\eqdef \min_{a\in A_{v_d}} \rho(v_d,a), \\
    \alpha(v_d) &\eqdef \arg\min_{a\in A_{v_d}} \rho(v_d,a),
\end{align}
where the latter is nonempty but might contain several elements.

We now suggest to measure $G$'s degree of {\em negligence} in choosing $a\in A_{v_d}$ at $v_d$
by the excess risk w.r.t.\ the minimum possible risk,
\begin{align}\label{eq:Drho}
    \Delta\rho(v_d,a) &\eqdef  \rho(v_d,a) - \underline\rho(v_d)
\end{align}
Comparing (\ref{eq:Drho}) with (\ref{eq:Dgamma}) and (\ref{eq:Dmu}), 
we see that this variant is still sensitive to all scenarios (like variant 3) 
rather than just the best-case (as in (\ref{eq:Dgamma})) or the worst-case (as in (\ref{eq:Dmu})). 
In particular, if a strategy $\sigma$ is weakly dominated by some undominated strategy $\sigma'$,
then using $\sigma$ is considered negligent even if the difference between $\sigma$ and $\sigma'$ only matters
in cases other than the best or worst.

Now, to measure $G$'s {\em backward responsibility} regarding $\eps$ in node $v_o\in V_o$,
we suggest to take their aggregate negligence over all choices taken,
\begin{align}
    \R_b^4(v_o) &\eqdef  \sum_{v_d\in H(v)\cap V_G} \Delta\rho(v_d,C_{v_d}(v_o)).
\end{align}
This now fulfils (NUR) again since 
by using a {\em risk-minimizing strategy} $\sigma$ for which $\sigma(v_d)\in\alpha(v_d)$ for all $v_d\in V_G$, 
$G$ can avoid all backward responsibility.

\paragraph{Maximum degree of negligence, forward responsibility.}
In analogy to variants 1 and 2, to measure $G$'s {\em forward responsibility} regarding $\eps$ in node $v_d\in V_G$,
we suggest to take the maximal possible degree of negligence,
\begin{align}
    \R_f^4(v_d) &\eqdef  \max_{a\in A_{v_d}} \Delta\rho(v_d,a) = \max_{a\in A_{v_d}} \rho(v_d,a) - \underline\rho(v_d).
\end{align}

We can now summarize some first results before turning to applying the above RFs in the social choice context.
\begin{proposition}\label{prop:main}
    Compliance of variants 0--4 with axioms (IND), (IAT), (GSM), (AMF), (NRV), (NUR), (MBF), (NFT), (NUD), (MFR), and (CFR) is as stated in Table \ref{tbl:axioms}.
\end{proposition}
\begin{table}
\footnotesize
\begin{tabular}{cccccccccccc} \toprule
\textbf{Variant} & (IND) & (IAT) & (GSM) & (AMF) & (NRV) & (NUR) & (MBF) & (NFT) & (NUD) & (MFR) & (CFR) \\ \midrule
0                & \cm   & ---   & \cm   & n/a   & \cm   & \cm   & n/a   & n/a   & n/a   & \cm   & --- \\
1                & \cm   & ---   & \cm   & ---   & \cm   & \cm   & \cm   & \cm   & ---   & \cm   & --- \\
2                & \cm   & \cm   & ---   & ---   & ---   & \cm   & \cm   & ---   & ---   & ---   & \cm \\
3                & ---   & \cm   & ---   & \cm   & \cm   & ---   & ---   & \cm   & \cm   & \cm   & \cm \\
4                & ---   & \cm   & ---   & ---   & ---   & \cm   & \cm   & \cm   & \cm   & \cm   & \cm \\ \bottomrule
\end{tabular}
\caption{\label{tbl:axioms}
Summary of selected axiom compliance by the suggested variants of $(\R_f,\R_b)$ 
}
\end{table}

\section{Application to social choice problems}\label{sec:socialchoice}

In this section, we apply the above-defined responsibility functions for
measuring degrees of forward and backward responsibility
to a number of social choice problems in which an electorate of $N$ voters
uses some election or decision method or social choice rule to choose exactly one 
out of a number of candidates or options, one of which, $U$, is ethically undesired.
We are interested in the forward responsibility of a group $G$ of $m$ voters to avoid the election of $U$ at each stage of the decision process,
and the backward responsibility of $G$ for $U$ being elected.

We first consider deterministic single-round decision methods in which all voters vote simultaneously
and probability plays a marginal role only to resolve ties,
and significantly probabilistic single-round decision methods.
Afterwards, we study a selection of two-round methods in which voters act twice with some sharing of information between the two rounds.
Finally, we turn to an example of a specific stylized social choice problem related to climate policy making.

We exploit all symmetry and independence properties 
heavily when modeling the otherwise rather large decision trees.
In particular, in each round, we implicitly treat a group $G$ of $m\ge 1$ many voters 
as equivalent to a single agent whose action set in a certain round consists of 
all possible combinations of $G$'s ballots.
We then model the simultaneous decision of all voters in a certain round 
by a single decision node for $G$,
followed
by ambiguity nodes representing the choices of the $m'\eqdef N-m$ many other voters,
one for each possible way or class of ways in which the members of 
$V-G$ 
might vote,
as exemplified in Fig.\ \ref{fig:majority}.

For simplicity, we do not discuss bordering cases in which ties may occur,
in particular by assuming the number of voters $N$ is odd.
Our results are summarized in Table \ref{tbl:vals}.

\subsection{Single-round methods}

\paragraph{Two-option majority voting.}

This is the simplest classical case.
Besides the ethically undesired option $U$, there is only one other, ethically acceptable option $A$,
and the event $\eps$ to be avoided is the election of $U$.
Each voter votes for either $U$ or $A$, with no abstentions allowed, and the option with more votes is elected.

We find that $\R_f^3(G)=1$ no matter how small $m$ is, since in the scenario where about half of the other voters vote for $U$, 
$G$'s voting determines whether $U$ is elected ($\ell=1$) or $A$ ($\ell=0$).
By contrast, $\R_f^2(G)=1$ only if $m > N/2$, otherwise $\R_f^2(G)=0$ since then $G$'s worst-case likelihood is always 1. 
Similarly, $\R_f^1(G)=1$ only if $m > N/2$ since only then they can guarantee a likelihood of 1.

Now assume $u$ voters from $G$ (and an arbitrary number of the other voters) have voted for $U$.
Obviously, $\R_b^0=\R_b^1(G)=1$ iff $u > N/2$, since only that guarantees a likelihood of 1.

To determine $\R_b^2(G)$, we notice that for $m < N/2$, $G$'s worst-case likelihood is always 1, 
so $G$ has zero degree of deviation and $\R_b^2(G)=0$; 
for $m > N/2$, the (unconditional) minimax likelihood is $\mu=0$; 
the conditional minimax likelihood given $u$ is $1$ if $m' > N/2 - u$, otherwise $0$.
This implies that $\R_b^2(G)=1$ only if $u > m-N/2 > 0$, otherwise $0$.

\begin{figure}
    \centering\includegraphics[width=.50\textwidth]{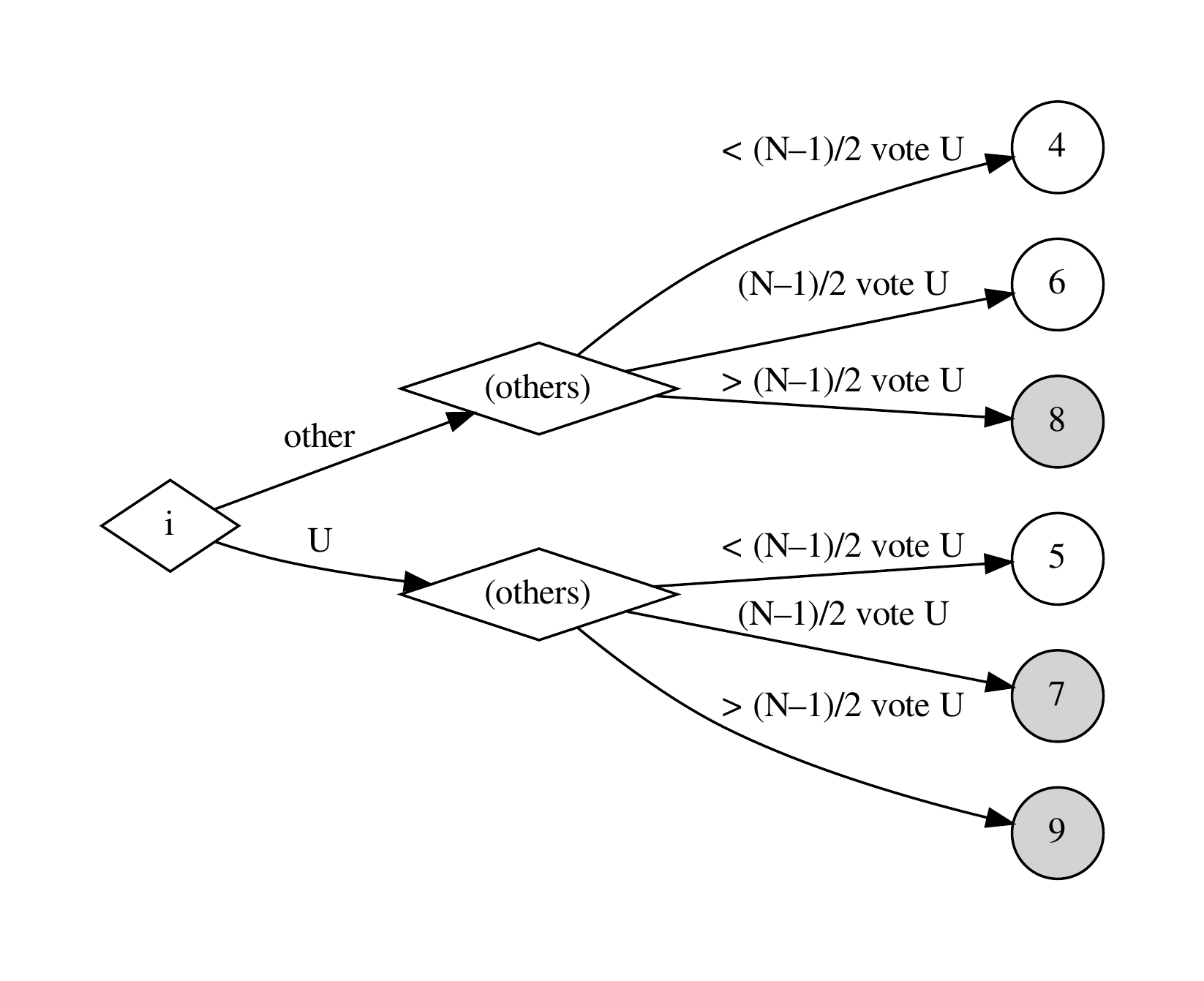}
\caption{\label{fig:majority}
Two-option majority voting from the perspective of a single voter $i$ who can either voter for the
ethically undesired option $U$ or another, acceptable option, not knowing how the other $N-1$ voters vote.}
\end{figure}

To determine $\R_b^3(G)$, we first consider a scenario $\zeta$ in which $v$ of the others have voted for $U$;
then $G$'s optimum is $\omega(\zeta)=1$ iff $v > N/2$, otherwise $\omega(\zeta)=0$,
and $G$'s optimum after choosing $u$ is $1$ iff $v > N/2 - u$, otherwise $0$;
hence $G$'s shortfall in $\zeta$ is $\Delta\omega(\zeta)=1$ iff $N/2 > v > N/2 - u$, otherwise $\Delta\omega(\zeta)=0$.
For $m=1$, the relevant distinction w.r.t.\ $v$ is depicted in Fig.\ \ref{fig:majority}.
So $G$'s risk taken by choosing $u$ was $\rho=1$ iff such a scenario exists,
i.e., iff $u>0$ and $m' > N/2-u$, otherwise $\rho=0$.
This implies that $\R_b^3(G)=1$ if $u > \max(m-N/2, 0)$, otherwise $\R_b^3(G)=0$.
Since putting $u=0$ is a weakly dominant strategy, $\R_b^4 = \R_b^3$ here.

By comparison, we see that in variants 3 and 4 of our responsibility functions, 
minorities can have nonzero forward and backward responsibility for the election outcome,
while in variant 2 only majorities can. 
In particular, under variants 3 and 4 every single voter who voted for $U$ has full backward responsibility 
since they took the risk that theirs would be the deciding vote.

Also, in all variants all degrees of responsibilities are either zero or one, 
and the actual voting behaviour of the others is irrelevant for the assessment of backward responsibility.

Note that this is in contrast to the ad-hoc idea that backward responsibility of $G$
should be a more smoothly increasing function of the number $u$ of voters from $G$ that voted for $U$
or maybe even proportional to $u$.

\paragraph{Random dictator.}

A major contrast is given by a method that is rarely used in practise 
but often used as a theoretical benchmark in social choice theory,
the ``random dictator'' method.
In addition to option $U$, there are any number of other, ethically acceptable options.
Each voter votes for one option, then a voter is drawn at random and their vote decides the election.

As $G$ controls exactly a share $m/N$ of the winning probability, 
their influence on $U$'s likelihood is $m/N$ in all scenarios, hence $\R_f^3=m/N$.
Also $\R_f^1=\R_f^2=m/N$ since their actions span a range of guaranteed likelihoods or worst-case likelihoods of width $|m/N|$.
When $u$ in $G$ have voted for $U$, their shortfall is $u/N$ in all scenarios, hence $\R_b^3=u/N$.
Also $\R_b^1=\R_b^2=u/N$ since their action increased the guaranteed or worst-case likelihood by $u/N$.

But $\R_b^0=0$ unless $u=m=N$ since for $u<N$ a positive probability for $\neg U$ remains.
This shows that in situations with considerable stochasticity, 
assessments based on deterministic causation such as $\R_b^0$ differentiate too little to be of any practical use.

\paragraph{Multi-option simple majority.}

Coming back to majority voting, we next study the case of more than two options, 
and will see that this leads to much more complicated analysis.
With an undesirable option $U$, $k\ge 2$ acceptable options $A_j$, and the possibility to abstain, 
we assume the winner is elected by lot from those that got the largest number of votes.

Suppose $u$ of the $m$ voters from $G$ vote for $U$ and $a_j$ for $A_j$,
with $\max_j a_j = a$.
Then the guaranteed likelihood of $U$ and the value of $\R_b^1$ are 0 (since the others can avoid $U$ for sure), iff $u-a<m'$, 
they are in $[1/(k+1),1/2]$ iff $u-a=m'$,
and they are 1 iff $u-a>m'$.
So $G$ can increase the guaranteed likelihood by 1 (i.e., $\R_f^1=1$) iff $m>\N2$ (since only then they can make both $u-a<m'$ and $u-a>m'$),
and otherwise (iff $m<\N2$), $\R_f^1=0$. 

Likewise, the worst-case likelihood of $U$ is 1 (since the others can make $U$ win for sure) iff $a-u<m'$, 
it is in $[1/(k+1),1/2]$ iff $a-u=m'$,
and it is 0 iff $a-u>m'$.
So $G$ can increase the minimax likelihood by 1 (i.e., $\R_f^2=1$) iff $m>\N2$ (since only then they can make both $a-u<m'$ and $a-u>m'$),
and otherwise (iff $m<\N2$), $\R_f^2=0$.
Hence $\R_b^2=1$ iff $m>\N2$ and $a-u<m'$,
$\R_b^2\in[1/(k+1),1/2]$ iff $m>\N2$ and $a-u=m'$,
and $\R_b^2=0$ otherwise.

If all others abstain, $G$'s influence is 1, hence $\R_f^3=1$ no matter how small $G$.
Assume $a-u<m'$. 
Then of the others, $v=a-u+1$ many could vote for $U$ and all others could abstain,
so that $U$ gets elected for sure.
If $a-u<m-1$, $G$ could then have avoided this outcome by increasing $a-u$ by 2.
Hence if $a-u<\min(m',m-1)$, $G$ takes risk 1 and gets $\R_b^3=1$.
But also if $u=0$, $a=m$, and $m'\ge m+3$, $G$ takes risk 1,
since of the others, $v=m+1$ many could voter for $u$, two for a third option $A_i$ that $G$ did not vote for, and all others could abstain,
so that again $U$ gets elected for sure and $G$ could have avoided it by voting for the same $A_i$ as the others.
This shows that for $m<\N2-1$, $\R_b^3=1$ no matter what $G$ does,
hence $\R_b^4=0$ no matter what $G$ does, and hence $\R_f^4=0$.

So, in contrast to the two-option case, in the multi-canditate case only variant 3 
assigns full backward responsibility to a single voter who votes for $U$,
while variant 4 acknowledges the possible excuse that, because there is not a unique contender to $U$,
and hence no weakly dominating strategy,
also any other way of voting of the single voter could have helped $U$ win.

But variant 3 has the major problem that it assigns full backward responsibility 
to every minority regardless of their behaviour.

\paragraph{Approval voting.}

Here everything is just as in multi-option simple majority, 
except that a voter can now vote for any number of options at the same time \cite{brams1978approval}.
Also the analysis is the same as before,
except that now also a minority $G$ has a weakly dominating strategy that reduces their risk to zero,
namely voting for all options but $U$.
As a consequence, now $\R^4=\R^3$ again,
and a single voter voting for $U$ but no other option has full backward responsibility in variants 3 and 4.

\paragraph{Full consensus or random dictator.}

As another probabilistic voting method, let us look at a method studied in \cite{Heitzig2010a} that was designed to give each group of voters an effective decision power proportional to their size (in contrast to majoritarian methods which give each majority full effective power and no minority any effective power).

In this method, each voter marks one option as ``favourite'' and one as ``consensus''.
If all mark the same option $X$ as consensus, $X$ is elected,
otherwise the option marked as favourite on a random ballot is elected.
Let $u$ be $G$'s ``favourite'' votes for $U$ and $a_j$ $G$'s ``consensus'' votes for option $A_j\neq U$.

If no member of $G$ distinguished between her two votes, the analysis is the same as for the random dictator method.
If all in $G$ put some $A\neq U$ as consensus ($a=m$),
the guaranteed likelihood of $U$ stays at zero since $A$ might still get 100\% winning probability.
In that case, $\R_b^1=0$ even if $U$ wins.

All worst-case likelihoods come from scenarios where the others specify $U$ as consensus, 
so the assessment in variant 2 is the same as in random dictator.

Regarding risk, however, we find that always $\rho=(u+m')/N$. 
This is because there is 
a scenario where everyone not in $G$ elects $U$ as favourite and the same option $A$ as consensus. In this case $G$'s optimal strategies are those where everyone also selects $A$ as consensus, leading to a zero probability of $U$ winning. If some of $G$'s members select another option as consensus, the resulting likelihood of $U$ being elected is $(u + m') / N$, which amounts to the risk taken by $G$. 

Hence $\R_b^3=(u+m')/N$ and $\R_f^3=1$.
Since the least possible risk is $\rho_0=(m')/N$,
$\R_b^4=\R_b^3-\rho_0=u/N$ and $\R_f^4=m/N$. 

Note that as $G$ cannot know which option the others select they cannot in the scenario described above know which option to select as consensus.

\paragraph{Other majoritarian methods.}

In any single-round method in which any group of $m>\N2$ many members 
have a way of voting that enforces any option they might choose,
we will have $\R_f^1=\R_f^2=1_{m>\N2}$ and $\R_f^3=1$.

\paragraph{Other proportional power allocating methods.}

In any single-round method in which any group of $m$ many members
have a way of voting that guarantees any option they might choose a probability of at least $m/N$,
we will have $\R_f^1=\R_f^2=m/N$,
$\R_b^0\le 1_{m=N}$.

\subsection{Two-round methods}

Real-world social choice situations often turn out to consist of several stages upon closer examination even when the ``main'' voting activity consists of all voters acting ``simulteneously''.
There are many ways in which decisions taken before or after the main voting stage may be relevant, including 
the pre-selection of options or candidates put on the menu e.g.\ via ``primaries'', 
taking and publishing any pre-election polls, 
seeing an election in the context of previous and future elections,
using one or several run-offs rounds to narrow down the final choice, 
challenging a decision afterwards in courts, etc.

We select here three paradigmatic examples that we believe cover most of the essential aspects:
(i) a round of pre-election polling as a very common example of ``cheap talk'' before the actual decision that has no formal influence on the result;
(ii) the possibility of amending an option as an example of influencing the menu of a decision that is very common in committee and parliamentary proceedings;
(iii) a simple runoff round after taking the main vote, as an example of an iterative procedure commonly used in public elections in order to ascertain a majority.

\paragraph{Simple majority with a pre-election poll.}

Before the actual voting by simple majority, 
a poll is performed and the options' total vote shares are published,
so voters might form beliefs about each others' eventual voting.
But since our responsibility measures are independent on any beliefs the agents might form 
about the probabilities of other agents' unknown choices at free will,
which are rather treated like any other ambiguity,
the polling has no influence on our assessment. 
Backward responsibility depends only on actual voting behaviour,
and forward responsibility is zero when answering the poll.
The same is true of any other form of pre-voting ``cheap talk''.

In particular, in all our variants, even the prediction of a landslide victory of $U$
does not reduce the responsibility to help avoiding $U$.

\paragraph{Two-option majority with an amendment round.}

Now we turn to a case where the repeated choices can really lead 
to changed responsibilities which might even exceed 1 in case of repeated failure to avoid $U$.

In round 1, $U$ is compared to an amended, ethically acceptable version $A$.
If $U$ wins, it is compared to another acceptable option $B$ in round 2.
Abstentions are not allowed.
Assume $a$ of the $m$ voted $A$ in round 1 and $b$ of the $m$ voted $B$ in round 2 
(if round 2 is not reached since $A$ won in round 1, then we put $b=m$).

$U$ can only be caused for sure or its guaranteed or worst-case likelihood increased in round 2
if $m>\N2$, by putting $b<\N2$;
hence in round 1, $\R_f^1=\R_f^2=0$; in round 2, $\R_f^1=\R_f^2=1_{m>\N2}$;
and eventually $\R_b^0=\R_b^1=\R_b^2=1_{b<\N2}$.
Also, $G$'s maximal influence is $\R_f^3=1$ in both rounds since their votes might make a difference. 

From the simple majority analysis, we know already that $\R_f^3=\R_f^4=1$ in round 2
and $\R_b^3,\R_b^4$ have a summand $1_{b<\min(\N2,m)}$ from their action in round 2.
Assume of the $m'$ others, $a'$ will vote $A$ in round 1 and $b'$ would vote $B$ in round 2.
In round 1, $G$'s optimum likelihood is then 0 iff they can prevent $U$, 
i.e., iff $\max(a',b')>\N2-m$, otherwise it is 1.
After voting in round 1, the optimum changes from 0 to 1 (so that $G$ has a shortfall of 1) 
iff $a+a'<\N2$ and $\max(a',b')>\N2-m$ but $b'<\N2-m$, i.e., iff $b'<\N2-m<a'<\N2-a$;
otherwise $G$'s shortfall is 0.
$G$'s risk taken by choosing $a$ is then $1$ iff the others can choose $a',b'$ so that $b'<\N2-m<a'<\N2-a$,
i.e., iff $a<m<\N2$.
Hence $\R_b^3$ has a summand $1_{a<m<\N2}$ from their action in round 1.
Since putting $a=b=m$ is a weakly dominating strategy, this implies that
$\R_b^3=\R_b^4 = 1_{a<m} + 1_{b<m}$ if $m<\N2$
and $\R_b^3=\R_b^4 = 1_{b<\N2}$ if $m>\N2$.

Note that all variants of $\R_b$ can exceed 1 here.

\paragraph{Simple runoff.}

After a simple vote on the three options $U,A,B$ without abstentions (round 1), 
either the one with an absolute majority (more than $\N2$ votes) wins
or $U$ has the fewest votes, or a round 2 is taken where $U$ is compared to the other front-runner.

Let $a,b$ be $G$'s votes for $A,B$ in round 1 and $u$ their votes for $U$ in round 2
(or 0 if there is no round 2).
For simplicity, we ignore ties here.

$U$ can be caused if $m>\N2$, by putting $a+b<\N2$ in round 1 or $u>\N2$ in round 2,
giving the values for variants 0--1 (see Table\ \ref{tbl:vals}), and again maximal influence is 1 in both rounds.
By choosing $a,b$ in round 1, $G$ increases minimax likelihood from 0 to 1 iff $a+b<\N2<m$.

Again, $\R_f^3=\R_f^4=1$ in round 2 and $\R_b^3,\R_b^4$ have a summand $1_{u>\max(m-\N2,0)}$ from round 2.
Assume the scenario in which the corresponding vote counts of the $m'$ others are $a',b',u'$.
In round 1, $G$'s optimum likelihood is then 0 
if they can either exclude $U$ in round 1 by putting $\min(a+a',b+b')>N-a-b-a'-b'$,
which is possible iff $a'+b' > 2N/3 - m$,
or if they can make $U$ lose in round 2, which is possible iff $u' < \N2$.
By choosing $a,b$, they increase the optimum likelihood to 1 
if they either make $U$ win in round 1 by putting $a + b < \N2 - a' - b'$
or if they let $U$ get to round 2 by putting $\min(a+a',b+b')<N-a-b-a'-b'$ 
in a situation where they cannot avoid that $U$ will win in round 2 since $u' > \N2$.
In all, their shortfall in round 1 will be 1 
iff ($a'+b' > 2N/3 - m$ or $u' < \N2$) and ($a + b < \N2 - a' - b'$ or ($\min(a+a',b+b')<N-a-b-a'-b'$ and $u' > \N2$)).
This is equivalent to 
$2N/3 - m < a'+b' < N/2 - a - b$
or
($a'+b' > 2N/3 - m$ and $\min(a+a',b+b')<N-a-b-a'-b'$ and $u' > \N2$)
or 
($u' < \N2$ and $a + b < \N2 - a' - b'$).
Such a scenario $(a',b',u')$ exists iff $a + b < \max(m - N/6, \N2)$,
so this is the condition for having taken a risk of 1 in round 1,
contributing to $\R_b^3$.
Since it can be avoided by putting $a+b=m$, $\R_b^4=\R_b^3$.

\subsection{Median voting for an emissions cap}

We finally analyze a stylized probabilistic example from global climate policy making
inspired by Weitzman's discussion of a fictitious World Climate Assembly \citep{Weitzman2017}.
Assume countries vote on a global greenhouse gas emissions cap (or, alternatively, a carbon price) by median voting.
Each country $i$ specifies an amount $a_i\ge 0$ of global emissions
and their median $\text{med}(a_1,\dots,a_N)$ is realized as a global cap.
This can be seen as a shortcut to taking a series of binary majority decisions 
in each of which the cap may be lowered or raised by some amount.

As a consequence of the resulting emissions-induced temperature increase, 
a certain climatic tipping element \citep{Lenton2008} may tip ($\eps$), 
leading to undesired economic damages and loss of life.
Let $f(a)$ be the best estimate of the probability of tipping given cap $a$,
based on the current state of scientific knowledge,
and assume $f(a)$ is weakly increasing in $a$, $f(c_0)=0$ and $f(c_1)=1$ for some values $c_0,c_1$.

Tipping can only be caused only if $m>\N2$, e.g. by putting $a_i\ge c_1$ for all $i\in G$.
Assuming the $m$ members of $G$ voted $a_1\le\cdots\le a_m$,
we hence get $\R_b^0 = 1$ if $m>\N2$ and $a_{m-(N-1)/2}\ge c_1$, else 0.
Similarly, $\R_b^1 = f(a_{m-(N-1)/2})$ if $m>\N2$, else 0;
hence $\R_f^1 = 1_{m>\N2}$.

The worst-case likelihood before voting is $1_{m<\N2}$.
After voting, it is $f(a_{(N+1)/2})$ if $m>\N2$, else 1.
Hence $\R_b^2 = f(a_{(N+1)/2})$ if $m>\N2$, else 0,
and also $\R_f^2 = 1_{m>\N2}$.

In the scenario where the others vote $a'_1\le\cdots\le a'_{m'}$,
$G$'s optimum likelihood before voting is $\omega = f(a'_{m'-(N-1)/2})$ if $m<\N2$, else 0.
Their shortfall is $\Delta\omega = \text{med}(a_1,\dots,a_N) - \omega$.
Given $G$'s votes, if $m>\N2$, the scenario that maximizes this shortfall is 
when all others vote $c_1$ so that $\Delta\omega = f(a_{(N+1)/2})$.
If $m<\N2$, it is when $(N+1)/2-m$ many of the others vote $c_0$ and the rest $c_1$
so that $\Delta\omega = f(a_m)$.
In all, $G$'s risk by voting $a_G$ is $\R_b^3 = f(a_{\min(m,(N+1)/2)})$.
This is minimized to 0 by the weakly dominant strategy of putting $a_G\equiv c_0$,
hence $\R_b^4=\R_b^3$, $\R_f^3=\R_f^4=1$.

\section{Discussion}\label{sec:discussion}

In this section we will present a discussion of a selection of responsibility ascriptions resulting from the application of our proposed functions to the paradigmatic examples presented in the beginning. This will include reference to certain of the desired axioms as well as to properties of existing formalisations, and the question of whether or not there are fulfilled by the corresponding functions. We will also discuss a selection of results of the application of the responsibility functions in the social choice scenarios.

\paragraph{(MFR), (CFR) and luck.} As was discussed above when introducing the paradigmatic example scenarios, we believe that in a situation where an agent did not know whether their action was going to have an effect, such as those represented in Figure~\ref{fig:paradigmatic}(a), node 4, and Figure~\ref{fig:paradigmatic}(b), node 6, the responsibility ascription should be made on the basis of their having to assume that their action was going to have an effect. The rationale behind this is precisely to disable dodging by referring to certain assumptions about the state of the world, since such assumptions would form an ``unreasonable belief'' in certainty in an actually uncertain situation. This relates to the discussion on {\em moral luck}, and the statement by \cite{vallentyne2008bruteluck} that we mentioned in the introduction, arguing for disregarding effects outside of the agents' control. these considerations are reflected in the axioms (MFR) and (CFR). 
$\R^1_b$ and $\R^2_b$ diverge on this question, with $\R^1_b$ assigning full backwards responsibility in the case of throwing a rock even though the window would have shattered anyways, but not assigning any responsibility for shooting when the gun was not loaded, and $\R^2_b$ giving inverse results. Ideally, both axioms would be fulfilled by a responsibility function. In variant 3, we managed this by basing it on a maximum over likelihood differences (which we call `risk' in this context) rather than on a difference of minimax likelihoods. Since this introduced some overdetermination, we further modified the formula in variant 4 to give agents again a way to avoid blame.





\paragraph{Voids.}
One important topic when talking about responsibility ascription in voting scenarios and other interactive settings is the potential of responsibility voids, as discussed in \cite{brahamvanhees2018}. While such situations, in which one does not assign responsibility to any single agent due to the interactions of several agents and/or nature, cannot occur in our variants 0, 1, and 3, they do exist in our variants 2 and 4. It is, however, our intuition that this is not a serious problem as long as one assigns responsiblity to at least some nonempty group of agents, which variants 2 and 4 do. Still, it makes these variants fail group subadditivity (see Appendix).

\paragraph{Effect of reducing ambiguity by learning.}


Consider Fig.\ \ref{fig:learn}, 
which is a stylized version of the decision situation humanity faced around the 1970's regarding climate change,
when it was already clear that humanity can influence global mean temperature via greenhouse gas emissions,
but when it was still unclear that there was a risk of undesired global warming 
rather than one of undesired global cooling due to the onset of glaciation.

Let us at first assume that the option to learn about which of the two scenario was correct was unavailable,
Fig.\ \ref{fig:learn}(a).
Then, in nodes 4 and 5, where humanity chooses to either heat-up the Earth (via high GHG emissions) or not (via low emissions), 
we have $\R_f^3=1$ since their choice would definitely make a difference,
but $\R_f^2=0$ since they don't know what the right choice is.
Likewise, in nodes 10 and 11 we have $\R_b^3=1$ and $\R_b^0=1$ ,
since then they made the wrong choice in node 4 or 5,
but $\R_b^2=0$ since they didn't know it was the wrong choice.
Even in nodes 9 and 12,
we then have $\R_b^3=1$ since even though their choice was right, it could have been wrong,
while $\R_b^2=0$ since their choice was not wrong from a worst-case avoidance perspective,
and $\R_b^0=0$ since their choice was right in the true scenario.

The fact that variant 2 does not assign responsibility in nodes 10 and 11 
since $G$ did not know they were causing the undesired event of large-scale climate change
might seem problematic since it might seem that such a method of assessing responsibility
gives perverse incentives to remain ignorant to avoid responsibility.

However, we will see that when we take into account that an agent has chosen to remain ignorant,
the responsibility assessment will reflect this.
Let us therefore now take into account the learning option in nodes 1 and 2 in Fig.\ \ref{fig:learn}(b).
While this does not change $\R_f^3$ and $\R_f^2$ in nodes 4 and 5,
it changes the remaining values.
In nodes 1 and 2, we get $\R_f^3=1$ since their future choices will make a difference,
and also $\R_f^2=1$ since the worst-case likelihood is 0 for the learning option but 1 for the passing option.
So learning is a minimax strategy here and consequently $\R_b^2$ counts the choice to pass as a deviation of degree 1,
now leading to $\R_b^2=1$ in nodes 10 and 11 (as well as nodes 8, 9, 12, and 13).
$\R_b^2=0$ only in nodes 7 and 14 where both choices made were correct.
$\R_b^3$ behaves the same as $\R_b^2$ here.
$\R_b^0$ however, caring only about causation, is unaffected by knowledge 
and so still has value 1 exactly in those nodes belonging to $\eps$: 8, 10, 11, 13.

Needless to say, today humanity is in node 3, where both variants 1 and 2 assign full forward responsibility,
even though climate science ``sceptics'' claim we're in information set $\{4,5\}$ or still in $\{1,2\}$, 
or even deny the whole model.

\paragraph{Effect of reducing ambiguity coordination.}

When we replace the initial ambiguity node of Fig.\ \ref{fig:learn}(a) and (b) by another agent $j$'s decision node, (a) becomes formally equivalent to a pure coordination game such as choosing one of two possible places to meet, and (b) can represent the possibility that $i$ can call $j$ to ask where $j$ will go, in order to coordinate. This means that also coordination can reduce responsibility. 
Likewise, in a social choice situation, voters who could coordinate their votes but fail to do so will be more responsible. 
The same will be true when voters could inform themselves about the likely consequences of the given options but fail to do so.

\paragraph{Relationship between individual and group responsibility.}

In \cite{braham2009degrees}, the example of three walkers together freeing a jogger from below a fallen tree is discussed, under the assumption that all three help lifting the trunk while two would have sufficed to lift it. In that example, which is formally equivalent to the two-option majority voting with three voters, if we assume neither walker sees whether the others are really lifting the trunk or just pretending to, our variants 0--2 judge no single walker but each pair of walkers (as well as all three together) as responsible for freeing the jogger. Variants 3 and 4 also judge each single walker as responsible.

The two-option majority example is also enlightening regarding the monotonicity demanded by (GSM).
As one can see from the dependence on $m$ shown in the top row of Table \ref{tbl:vals}, variants 2--4 fail (GSM) since they may assign a group less responsibility than its members in cases where one member's actions makes harmless the possibly bad consequences of another member's action who could however not trust that he will be this lucky. 
In such cases, these variants judge the latter member responsible, thereby fulfilling (CFR), but not the group.
Indeed it seems difficult to fulfill both (GSM) and (CFR) without also assigning a group responsibility in cases where they were not simply lucky.

Another type of relationship between a group and a group member appears in \cite{list2011group}. In this example, a group orders one member to do something who can then decide to follow the order or not, but who will not act without the order. Since the order does not guarantee any positive probability of action, our variants 0 and 1 will only hold the member responsible. Since it increases the worst-case probability and is risky, variants 2--4 will also hold the group responsible, which seems to conform better to the general intuition.

\paragraph{Influence of timing.}

As we discussed when arguing for the use of extensive-form games rather than the more common normal-form games, actions in real life hardly ever happen at the exact same time. Considering for example a situation with two agents, their actions can be regarded as practically coincidental if they do not know about the other's choice before deciding on their own action. This is represented in our model using information sets. However, according to the argument given here, it should not matter whether we represent the situation as one agent acting first and the other following up, or the other way around. However, in variants $\R^1_b$ and $\R^2_b$, this change of representation would counter-intuitively shift responsibility assignments between the agents. Variants $\R^3$ and $\R^4$ manage to avoid this fallacy by taking into account the agent's information set.

\section{Conclusions and Outlook}\label{sec:conclusion}






We have established that determining a representation of degrees of responsibility within interactive scenarios playing out over time with probabilistic uncertainty as well as ambiguity is an important endeavour.
Specifically in light of the current climate crisis it becomes of a special interest to provide calculations applicable to a set of election scenarios that take into account all of these complexities.

Certain existing calculations are of an ad hoc nature, reducing the issue of responsibility for climate change to cumulative past emissions or population shares. Others are of a foundational nature, providing representations for the concept of responsibility in general, but not considering all of the complications at the same time or distributing responsibility rather cautiously, leading to voids. 

In the present paper we followed the second route, by providing an account that applies to questions of responsibility in general, but specifically aims at accounting for complexities in real-world applications. We suggested a number of responsibility functions extending those previously presented in the literature, to open the discussion on available representations. 
We used an axiomatic method, as it is known from social choice theory, to evaluate the proposed functions in a rigorous way. 

\paragraph{Framework.}
The framework used in this paper is an extension of extensive-form games. This game-form includes a temporal aspect, allowing for agents to make choices successively. We used specific nodes to represent ambiguity and probabilistic uncertainty with respect to the state of the world after this node, and an equivalence relation to express agents' perception (equivalent nodes cannot be distinguished). As we do not apply game theoretic analyses, such as evaluations of strategies of rational agents, we did not require individual utility functions. Instead, a universal `ethical desirability' assessment was made, selecting a subset of the outcome nodes as undesirable.

\paragraph{Axioms.}
Having introduced the framework we presented some potentially desirable properties for prospective responsibility functions. Clearly, a large number of such properties come to mind. The ones that we decided to present in detail were mainly ones that differentiate between the proposed responsibility functions. One important feature that we kept in mind was that we wanted to avoid dodging of responsibility. That is, we wished to reduce the number of situations where an undesirable outcome occurs or can occur but an agent potentially involved in its bringing about can claim to have no responsibility.

The first desirable properties that we presented were a set of independence axioms regarding the specific representation of the decision scenario. That is, if one and the same situation can be represented using slightly different game trees, this should not influence the resultant responsibility assignments. While these properties are certainly desirable, they are not trivial, as it is well known that in a formalisation the specific choice of representation can have repercussions on the outcome (consider the `Queen of England' example from \cite{beebee2004} repeated in \cite{brahamvanhees2012}).

The next set of desirable properties were monotonicity requirements: first with respect to increasing group size, and second with respect to increasing knowledge.
               
Subsequently we included another set of very intuitive considerations that have already been discussed in the literature. These conditions are the avoidance of responsibility voids (in the absence of uncertainty \emph{someone} must be responsible for an undesirable outcome) and the possibility to avoid responsibility by following some strategy that is optimal in this respect. 

Next, an axiom to relate forward and backward responsibility ascription was introduced: the degree of forward responsibility of a group is bounded by their maximal degree of backward responsibility. 

Lastly, we presented a set of axioms relating to situations where the agent is unsure about the actual state of the world and they do not know whether they are in a position to have any effect at all on the outcome. One can argue that they need to take into account the possibility of their action being significant, and accept the corresponding responsibility.

\paragraph{Candidate responsibility functions.}
This set of axioms allowed for a fruitful comparison of several candidate responsibility functions. As the benchmark variant ($\R_{f/b}^0$) we studied a representation of `strict causation' in our framework: ascription of full backwards-looking responsibility to a group if and only if there was a specific node at which an agent from the group took a decision determining the undesired outcome. 
Clearly for this variant does no axiom relating to forward-looking responsibility applies, nor does it include a sensible idea of \emph{degrees} of responsibility. More importantly, however, it assigns zero responsibility if the agent's action, unbeknownst to the agent, was not actually going to affect the outcome. This, as we stated above, was something we wanted to avoid.

The next suggestion ($\R_{f/b}^1$) was a function that extended the idea of `strict causation' to a probabilistic context: we assigned responsibility to the degree that an agent has caused an increase in the guaranteed likelihood of the undesired event. Group responsibility arises as an aggregate of its members' responsibility. While this function does include a notion of degree, it may still assign no responsibility if there was uncertainty regarding the effects of an action. 

The previous function preferred ambiguity over probabilistic uncertainty when trying to avoid responsibility, as the goal then has to be to avoid increasing \emph{guaranteed} likelihood of the undesired outcome. As no probabilities are available for ambiguity nodes, the guaranteed likelihood of each of its successors remains zero. On the contrary, one can consider a responsibility function which acts in the opposite direction: preferring probabilistic uncertainty over ambiguity (as is the case for most human decision makers). We achieved this by considering increases in the minimal \emph{worst-case} likelihood of the undesired event to lead to responsibility ascription ($\R_{f/b}^2$). The prescribed action rationale behind this function can be seen as `avoiding the worst' (in order to avoid carrying responsibility), rather than optimising for the best, similar to the game-theoretic notion of maximin strategies. In situations where an agent does not know whether their action leads to an undesirable outcome or not this seems to be a reasonable consideration.
Surprisingly, however, this function may lead to responsibility voids.

Both of the above functions 
reduce the conceptual complexity of responsibility ascription by relying on comparisons between the action taken at a specific node and a `baseline case'. In order to exploit the full weight of the extensive-form game structure that we have at hand, we decided to refer to the information sets as well as to \emph{strategies}, i.e., full plans of action including reactions to future outcomes of other agent's choices or uncertainty resolutions. By employing these two features we managed to escape the responsibility voids due to uncertainty that we experienced with the previous function.

In the next variant ($\R_{f/b}^3$) we assigned responsibility whenever a hypothetical minimum was not reached. This avoided responsibility voids, however, it also resulted in groups sometimes being assigned responsibility no matter what their action was. In the final variant we therefore set their best option as the baseline to be compared to ($\R_{f/b}^4$). This avoided certain situations in which a group is always to some extent responsible, but it re-introduced voids that were absent with the preceding function.

As a bottom-line, even though we managed to fulfil all of the desired properties by one or another variant of responsibility functions, none of the functions studied so far complies to all of them.

\paragraph{Social choice.} 
In a next step we determined the responsibility ascription the proposed functions offer in specific social choice settings. The first set of methods we examined were single-round methods, starting with simple two-option majority voting. In line with our considerations when suggesting the different responsibility functions, it does not actually matter for responsibility ascription (in either function) what the others voted. Also, considering that majority voting contains no probabilistic component out of the proposed functions only $\R_f^3$ and $\R_f^4$ can assign non-zero forward responsibility to non-majority groups. This consideration was one of the reasons we introduced these measures, as they represent a very common intuition: even if a minority group (say one voter) cannot influence the outcome with certainty, they still carry a responsibility to avoid the undesirable candidate. 

As was to be expected, the benchmark responsibility function based on deterministic causation does not represent intuitions very well when we look at voting mechanisms that make use of probabilistic procedures. 

Multi-option majority voting is somewhat more complicated than the single-option case, and, notably, differentiates between responsibility functions $\R^3$ and $\R^4$. In the case of several alternatives with one of them being undesirable, it is not clear which other option to vote for in order to avoid the election of the undesired candidate, as any of the other candidates might turn out to be the strongest opponent. Our `strictest' function $\R_b^3$ does not allow for this excuse, but assigns full responsibility to every minority group, regardless of their behaviour. $R_b^4$ does allow for the described excuse and omits the unavoidable responsibility of minorities.

As a second set of voting methods we examined two-round methods, such as voting with a pre-election poll or simple runoff (between the two preferred candidates from the first round). As we explicitly did not consider assumptions about other's behaviour to have an influence on responsibility ascription, polls or any other means for forming beliefs about other's voting behaviour have no influence, in neither of our functions. If the ethically undesirable candidate is elected in a runoff scenario, a group's responsibility can rise above 1. 

As a last voting scenario, and to get back to our initial application of climate policy, we examined responsibility ascriptions in a hypothetical situation of median voting for an emissions cap (or carbon price). The unwanted tipping of a certain element is induced with a certain probability, depending on the elected cap. This example neatly represents the direct reflection of the probability measures in the responsibility ascription. $\R_b^0$, $\R_b^1$ and $\R_b^2$ assign zero probability to minority groups, as their votes can guarantee neither a positive nor a below-one probability of tipping. In contrast, variants $\R_b^3$ and $\R_b^4$ assign a minority group a responsibility that equals that tipping probability which corresponds to the largest cap that any of the group members suggested. In particular, a single voter suggesting some value of the cap is responsible to the exact degree that this cap would make tipping likely.

\paragraph{Outlook.}
Due to the integration of methods from different disciplines several paths to continue the work presented here offer themselves.


First of all, to give full credit to the axiomatic method employed here, it would be natural to determine logical implications or exclusions between the axioms, as well as a characterisation of certain groups of responsibility functions with respect to sets of axioms.


An alternative account of causation using a variant of the NESS test, which has a direct representation in our formalism, could be compared to the resulting functions developed here. 


The differences between variants 0 and 1 on one hand and 2--4 on the other suggest looking for compromise variants, either by taking a closer look at the literature on choice under ambiguity \citep{Etner2012}, or by combining several functions into one, e.g.\ $\R'=(\R^1+\R^2)/2$ etc.


As we put a great emphasis on applicability one should consider the computational complexity of the functions proposed here. As our calculations often employ minimal or maximal likelihoods of the undesired outcome the calculation will quickly require many steps if we take into account every action by every agent. However, employing means of reducing the underlying tree, followed by heuristics concerning what the worst and best scenarios will be, should allow for the application of these functions to actual real world decision problems currently at hand.


\bibliographystyle{plain}
\bibliography{extracted,jobst,sarah}

\begin{thebibliography}{10}

\bibitem{Andre1983}
Judith Andre.
\newblock {Nagel, Williams, and moral luck}.
\newblock {\em Analysis}, 43:202--207, 1983.

\bibitem{arrow2013should}
Kenneth~J Arrow, Maureen Cropper, Christian Gollier, Ben Groom, Geoffrey~M
  Heal, Richard~G Newell, William~D Nordhaus, Robert~S Pindyck, William~A
  Pizer, Paul Portney, et~al.
\newblock How should benefits and costs be discounted in an intergenerational
  context? the views of an expert panel.
\newblock {\em The views of an expert panel (December 19, 2013). Resources for
  the future discussion paper}, (12-53), 2013.

\bibitem{baron2016justification}
Marcia Baron.
\newblock Justification, excuse, and the exculpatory power of ignorance.
\newblock In {\em Perspectives on Ignorance from Moral and Social Philosophy},
  pages 65--88. Routledge, 2016.

\bibitem{beebee2004}
Helen Beebee.
\newblock {\em Causation and {C}ounterfactuals}, chapter Causing and
  {N}othingness, pages 291 -- 308.
\newblock MIT Press, 2004.

\bibitem{stit2001}
Nuel Belnap, Michael Perloff, and Ming Xu.
\newblock {\em Facing the {F}uture. {A}gents and {C}hoices in our
  {I}ndeterminist {W}orld}.
\newblock Oxford University Press, 2001.

\bibitem{botzen2008}
W.~J.~W. Botzen, J.~M. Gowdy, and J.~C. J.~M. van~den Bergh.
\newblock Cumulative {CO}$_2$ emissions: shifting international responsibilites
  for climate debt.
\newblock {\em Climate Policy}, pages 569 -- 576, 2008.

\bibitem{braham2009degrees}
Matthew Braham and Martin Van~Hees.
\newblock Degrees of causation.
\newblock {\em Erkenntnis}, 71(3):323--344, 2009.

\bibitem{brahamvanhees2012}
Matthew Braham and Martin van Hees.
\newblock An {A}natomy of {M}oral {R}esponsibility.
\newblock {\em Mind}, 121(483):601 -- 634, July 2012.

\bibitem{brahamvanhees2018}
Matthew Braham and Martin van Hees.
\newblock Voids or {F}ragmentation: {M}oral {R}esponsibility for {C}ollective
  {O}utcomes.
\newblock {\em The Economic Journal}, 128(612), 2018.

\bibitem{brams1978approval}
Steven~J Brams and Peter~C Fishburn.
\newblock Approval voting.
\newblock {\em American Political Science Review}, 72(3):831--847, 1978.

\bibitem{broersen2009}
Jan Broersen.
\newblock A stit-{L}ogic for {E}xtensive {F}orm {G}roup {S}trategies.
\newblock In {\em IEEE/WIC/ACM International Conference on Web Intelligence and
  Intelligent Agent Technology - Workshops}, 2009.

\bibitem{broersen2011mensrea}
Jan Broersen.
\newblock Deontic epistemic stit logic distinguishing modes of mens rea.
\newblock {\em Journal of Applied Logic}, 9:137 -- 152, 2011.

\bibitem{chocklerhalpern2004}
Hana Chockler and Joseph~Y. Halpern.
\newblock Responsibility and {B}lame: {A} {S}tructural-{M}odel {A}pproach.
\newblock {\em Journal of Artificial Intelligence Research}, 22:93 -- 115,
  2004.

\bibitem{guardian2011responsibility}
Duncan Clark.
\newblock Which nations are most responsible for climate change?
\newblock {\em The Guardian}, 2011.

\bibitem{Decerf2019}
Benoit Decerf and Frank Riedel.
\newblock {Purification and disambiguation of Ellsberg equilibria}.
\newblock {\em Economic Theory}, 2019.

\bibitem{duijf2018}
Hein Duijf.
\newblock {\em Let's {D}o {I}t! {C}ollective {R}esponsibility, {J}oint
  {A}ction, and {P}articipation}.
\newblock PhD thesis, Universiteit Utrecht, 2018.

\bibitem{ellsberg1961}
Daniel Ellsberg.
\newblock Risk, {A}mbiguity, and the {S}avage {A}xioms.
\newblock {\em The Quarterly Journal of Economics}, 75(4):643 -- 669, 1961.

\bibitem{Etner2012}
Johanna Etner, Meglena Jeleva, and Jean~Marc Tallon.
\newblock {Decision theory under ambiguity}.
\newblock {\em Journal of Economic Surveys}, 26(2):234--270, 2012.

\bibitem{gardiner2004}
Stephen~M. Gardiner.
\newblock Ethics and global climate change.
\newblock {\em Ethics}, 114:555 -- 600, April 2004.

\bibitem{Heitzig2010a}
Jobst Heitzig and Forest~W. Simmons.
\newblock {Some chance for consensus: Voting methods for which consensus is an
  equilibrium}.
\newblock {\em Social Choice and Welfare}, 38(1):43--57, nov 2012.

\bibitem{horty2001}
John~F. Horty.
\newblock {\em Agency and {D}eontic {L}ogic}.
\newblock Oxford University Press, 2001.

\bibitem{huckabee2007}
Mike Huckabee.
\newblock 2007 {GOP} primary debate, {S}imi {V}alley, {C}alifornia, 3 May 2007,
  \href{https://www.c-span.org/video/?197893-1/republican-presidential-candidates-debate}{www.c-span.org/video/?197893-1/republican-presidential-candidates-debate}.

\bibitem{kadish1985complicity}
Sanford~H Kadish.
\newblock Complicity, cause and blame: A study in the interpretation of
  doctrine.
\newblock {\em California Law Review}, 73:323, 1985.

\bibitem{kriegler2009imprecise}
Elmar Kriegler, Jim~W Hall, Hermann Held, Richard Dawson, and Hans~Joachim
  Schellnhuber.
\newblock Imprecise probability assessment of tipping points in the climate
  system.
\newblock {\em Proceedings of the national Academy of Sciences},
  106(13):5041--5046, 2009.

\bibitem{Lenton2008}
Timothy~M Lenton, Hermann Held, Elmar Kriegler, Jim~W Hall, Wolfgang Lucht,
  Stefan Rahmstorf, and Hans~Joachim Schellnhuber.
\newblock {Tipping elements in the Earth's climate system}.
\newblock {\em Proceedings of the National Academy of Sciences of the United
  States of America}, 105:1786--1793, 2008.

\bibitem{lenton2019tippingpoints}
Timothy~M. Lenton, Johan Rockstr\"om, Owen Gaffney, Stefan Rahmstorf, Katherine
  Richardson, Will Steffen, and Hans~Joachim Schellnhuber.
\newblock Climate tipping points -- too risky to bet against.
\newblock {\em Nature}, 575:592 -- 595, November 2019.

\bibitem{list2011group}
Christian List, Philip Pettit, et~al.
\newblock {\em Group agency: The possibility, design, and status of corporate
  agents}.
\newblock Oxford University Press, 2011.

\bibitem{ipcc1.5spm}
Masson-Delmotte, V., P.~Zhai, H.-O. P{\"o}rtner, D.~Roberts, J.~Skea, P.R.
  Shukla, A.~Pirani, W.~Moufouma-Okia, C.~P{\'e}an, R.~Pidcock, S.~Connors,
  J.B.R. Matthews, Y.~Chen, X.~Zhou, M.I. Gomis, E.~Lonnoy, T.~Maycock,
  M.~Tignor, and T.~Waterfield, editors.
\newblock {\em {IPCC}, 2018: {S}ummary for {P}olicymakers}.
\newblock 2018.

\bibitem{ipcc_uncertain}
Michael~D Mastrandrea, Katharine~J Mach, Gian-Kasper Plattner, Ottmar
  Edenhofer, Thomas~F Stocker, Christopher~B Field, Kristie~L Ebi, and
  Patrick~R Matschoss.
\newblock The {IPCC AR5} guidance note on consistent treatment of
  uncertainties: a common approach across the working groups.
\newblock {\em Climatic Change}, 108(4):675, 2011.

\bibitem{mueller2009}
Benito M{\"u}ller, Niklas H{\"o}hne, and Christian Ellermann.
\newblock Differentiating (historic) responsibilities for climate change.
\newblock {\em Climate Policy}, 9:593--611, 01 2009.

\bibitem{Nagel1979}
Thomas Nagel.
\newblock {Moral Luck}.
\newblock In {\em Mortal Questions}. Cambridge University Press, 1979.

\bibitem{nelkin2007}
Dana~K. Nelkin.
\newblock Do {W}e {H}ave a {C}oherent {S}et of {I}ntuitions about {M}oral
  {R}esponsibility?
\newblock {\em Midwest Studies in Philosophy}, XXXI:243 -- 259, 2007.

\bibitem{nordhaus2007b}
William Nordhaus.
\newblock A {R}eview of the {S}tern {R}eview on the {E}conomics of {C}limate
  {C}hange.
\newblock {\em Journal of Economic Literature}, 45:686 -- 702, 2007.

\bibitem{poetter2019}
Bernhard Poetter.
\newblock Eine {M}illiarde {T}onnen zu viel.
\newblock {\em taz. Die Tageszeitung}, 2019.

\bibitem{world2014turn}
World~Bank Publications.
\newblock {\em Turn down the heat: confronting the new climate normal}.
\newblock World Bank Publications, 2014.

\bibitem{ringius2002burden}
Lasse Ringius, Asbj{\o}rn Torvanger, and Arild Underdal.
\newblock Burden sharing and fairness principles in international climate
  policy.
\newblock {\em International Environmental Agreements}, 2(1):1--22, 2002.

\bibitem{schellnhuberrahmstorf2016}
Hans~Joachim Schellnhuber, Stefan Rahmstorf, and Ricarda Winkelmann.
\newblock Why the right climate target was agreed in {P}aris.
\newblock {\em Nature Climate Change}, 6:649 -- 653, 2016.

\bibitem{stern2010}
Nicholas Stern.
\newblock The {E}conomics of {C}limate {C}hange.
\newblock In Stephen~M. Gardiner, Simon Caney, Dale Jamieson, and Henry Shue,
  editors, {\em Climate Ethics. Essential readings}. Oxford University Press,
  2010.

\bibitem{tamminga2019irreducibility}
Allard Tamminga and Frank Hindriks.
\newblock The irreducibility of collectivce obligations.
\newblock {\em Philosophical Studies}, pages 1 -- 25, 2019.

\bibitem{axiomaticmethod}
William Thomson.
\newblock On the axiomatic method and its recent applications to game theory
  and resource allocation.
\newblock {\em Social Choice and Welfare}, (18):327 -- 386, 2001.

\bibitem{Tong2004}
R~Tong.
\newblock {Review: Risk and Luck in Medical Ethics by D. Dickenson}.
\newblock {\em J Med Ethics}, 2004.

\bibitem{vallentyne2008bruteluck}
Peter Vallentyne.
\newblock Brute {L}uck and {R}esponsibility.
\newblock {\em Politics, Philosophy and Economics}, 7(1):57 -- 80, 2008.

\bibitem{vincent2011}
Nicole~A. Vincent.
\newblock {\em Moral {R}esponsibility. {B}eyond {F}ree {W}ill and
  {D}eterminism}, chapter~2, pages 15 -- 35.
\newblock Springer, 2011.

\bibitem{Weitzman2017}
Martin~L. Weitzman.
\newblock {Voting on prices vs. voting on quantities in a World Climate
  Assembly}.
\newblock {\em Research in Economics}, 71(2):199--211, 2017.

\bibitem{wunderlingetal2019}
N.~Wunderling, J.F. Donges, J.~Kurths, and R.~Winkelmann.
\newblock Interacting tipping elements increase risk of climate domino effects.
\newblock In review, 2019.

\bibitem{yazdanpanahdastani2016}
Vahid Yazdanpanah and Mehdi Dastani.
\newblock Quantified group responsibility in multi-agent systems.
\newblock In Corrado Santoro, Fabrizio Messina, and Massimiliano {De
  Benedetti}, editors, {\em Proceedings of the 17th Workshop {"}From Objects to
  Agents{"}}, CEUR Workshop Proceedings, pages 44--49, Italy, 2016. University
  of Catania.

\bibitem{zimmerman2016ignorance}
Michael~J Zimmerman.
\newblock Ignorance as a moral excuse.
\newblock In {\em Perspectives on ignorance from moral and social philosophy},
  pages 89--106. Routledge, 2016.

\end{thebibliography}

\section*{Appendix}

\subsection*{Longer list of axioms}

In this section, we compile a longer list of axioms which we believe may be relevant for the design of plausible RFs.
Similar to axioms in other branches of social choice theory, 
most of the axioms state that the value of an RF should not change or should change in a certain direction when some of its arguments
$\T,v,G,\eps$ are changed in certain simple ways.
We group the axioms roughly into categories, beginning with basic symmetry and independence axioms,
then listing certain possible monotonicity properties, 
and finally some that suggest certain values in specific situations.
We do not mean to suggest that all these axioms should be fulfilled,
only that there may be reasonable arguments why one might think it plausible to desire them.

\paragraph{Symmetry axioms, independence, and simplification axioms.}
Our first four axioms are similar to social choice theory's {\em anonymity} and {\em neutrality} axioms.
\begin{description}
\item[(Anon)] {\em Anonymity.} 
  If every occurrence of a certain individual $i\in I$ is replaced in both $\T$ and $(V,E)$ by a new individual $i'\notin I$, 
  then $\R(G)$ remains unchanged. 
  (I.e., individuals' identities are irrelevant beyond their influence on the outcome.)
\item[(ACon)] {\em Action-Related Consequentialism.} 
  If for some $v\in V_d$, a certain action $a\in A_v$ is replaced by a new action $a'\notin A_v$ in both $A_{v'}$ and $c_{v'}$ for all $v'\sim v$, 
  then $\R(G)$ remains unchanged. 
  (I.e., actions are only relevant via their (potential or actual) consequences.)
\item[(OCon)] {\em Outcome-Related Consequentialism.}
  If a certain outcome $v_o\in V_o - v$ is replaced in both $\T$ and $\eps$ by a new outcome $v_o'\notin V$,
  then $\R(v)$ remains unchanged;
  and if $v\in V_o$ and it is replaced in both $\T$ and $\eps$ by a new outcome $v_o'\notin V$,
  then the new $\R(v_o')$ equals the old $\R(v)$.
  (I.e., outcomes are only relevant via their belonging to $\eps$.)
\item[(FCS)] {\em Forward Complementation Symmetry.}
  If $\eps$ is replaced by its complement $\eps'=V_o-\eps$, forward responsibility does not change:
  $\R_f(\eps') = \R_f(\eps)$.
  (I.e., forward responsibility is about $G$'s influence on which of the two mutually exclusive events $\eps,\eps'$ obtains,
  not about which of the two is ethically desirable.)
\end{description}
Note that (ACon) in particular implies that there is no inherent difference between ``doing something'' and ``doing nothing''
despite their consequences.

The next five allow us to trim and coarse-grain a tree in certain ways, 
and are ruling out RFs that are based on some form of merely ``counting possibilities'':
\begin{description}
\item[(IST)] {\em Independence of Sure Thing Nodes.}
  If a node $v\in V-V_d$ or complete-information node $v\in V_d$ has only one successor, $S_{v}=\{v'\}$, it may be eliminated
  and replaced by $v'$ in $S_{P(v)}$ and $c_{P(v)}$.
  (I.e., nodes with only one possible successor are irrelevant for responsibility assessments.)
\item[(IZP)] {\em Independence of Zero Probabilities.}
  If a successor $v'\in S_{v_p}$ of a probability node $v_p\in V_p$ has zero probability, $p_{v_p}(v')=0$,
  then $v'$ and its branch may be ignored in assessing $\R(v)$ for any $v$ that is not contained in the branch $B(v')$.
  (I.e., possibilities that are ``almost surely'' not occurring are irrelevant for responsibility assessments.)
\item[(ICP)] {\em Independence of Cloned Possibilities.}
  Let $v_a\in V_a$ be an ambiguity node and $v'\in S_{v_a}$ one of its successors.
  Assume we add to $S_{v_a}$ another node $v''$ which is an exact copy of $v'$, followed by a branch $B(v'')$
  that is an exact copy of $B(v')$.
  Then $\R(v)$ must not change.
  (I.e., two identical possibilities are equivalent to just one copy of this possibility.)
\item[(INA)] {\em Independence of Nested Ambiguities.}
  If an ambiguity node $v_a\in V_a$ is succeeded by another ambiguity node $v_a'\in V_a\cap S_{v_a}$, 
  $v_a'$ may be ``pulled back'' into $v_a$, i.e., $v_a'$ may be eliminated and $S_{v_a'}$ added to $S_{v_a}$.  
\item[(INP)] {\em Independence of Nested Probabilities.}
  If a probability node $v_p\in V_p$ is succeeded by another probability node $v_p'\in V_p\cap S_{v_p}$, 
  $v_p'$ may be pulled back into $v_p$, i.e., $v_p'$ may be eliminated, $S_{v_p'}$ added to $S_{v_p}$, 
  and $p_{v_p}$ extended to $S_{v_p'}$ via $p_{v_p}(v'') = p_{v_p}(v_p')p_{v_p'}(v'')$ for all $v''\in S_{v_p'}$.
\end{description}
The next one is in a similar spirit but likely more debatable (see main text):
\begin{description}
\item[(IND)] {\em Independence of Nested Decisions.}
  If a complete-information decision node $v_d\in V_i$ is succeeded via some action $a\in A_{v_d}$ 
  by another complete-information decision node $v_d'=c_{v_d}(a)\in V_i$ of the same agent $i$, 
  then $v_d'$ may be pulled back into $v_d$, i.e., $v_d'$ may be eliminated, $S_{v_d'}$ added to $S_{v_d}$, 
  $\{a\}\times A_{v_d'}$ added to $A_{v_d}$, and $c_{v_d}$ extended by $c_{v_d}(a,a') = c_{v_d'}(a')$ for all $a'\in A_{v_d'}$.
\end{description}
Also more debatable are the following axioms which can be seen as treating the relationship of RFs to certain game-theoretic concepts.
The first one basically states that like in most equilibrium concepts for extensive-form games,
ambiguity and information-equivalence play a kind of complementary role:
\begin{description}
\item[(IAT)] {\em Independence of Ambiguity Timing.}
  Assume some probability node $v\in V_p$ or complete-information decision node $v\in V_d$ 
  is succeeded by an ambiguity node $v_a\in V_a\cap S_{v}$.
  Let $B(v)$, $B(v_a), B(v')$ be the original branches of the tree $(V,E)$ starting at $v$, $v_a$ and any $v'\in S_{v_a}$.
  For each $v'\in S_{v_a}$, 
  let $B'(v')$ be a new copy of the original $B(v)$ in which the subbranch $B(v_a)$ is replaced by a copy of $B(v')$;
  let $f(v')$ be that copy of $v$ that serves as the root of this new branch $B'(v')$.
  If $v\in V_d$, put $f(v')\sim f(v'')$ for all $v',v''\in S_{v_a}$
  Let $B'(v_a)$ be a new branch starting with $v_a$ and then splitting into all these new branches $B'(v')$.
  Then $v_a$ may be ``pulled before'' $v$ by replacing the original $B(v)$ by the new $B'(v_a)$.
\end{description}
The second one states that, in contrast to the most common game theoretic approach 
where players' ``optimal'' behaviour depends on the subjective probabilities they attach to others' behaviours,
for normative responsibility assessments only others' {\em possible} actions should play a role, not the agent's beliefs about their likelihoods;
as a consequence, other agents' actions could be seen as just another source of ambiguity:
\begin{description}
\item[(IOA)] {\em Independence of Others' Agency.}
  If $i\in I - G$, and some of $i$'s decision nodes $v_d\in V_i$ is replaced in $\T$ by a new ambiguity node $v_a\notin V$ with $S_{v_a}=S_{v_d}$,
  then $\R(G)$ remains unchanged 
  (i.e., it is irrelevant whether uncertain consequences are due to choices of other agents 
  or some non-agent mechanism with ambiguous consequences).
\end{description}
The third one is related to the branch of game-theory that studies group strategies and group deviations
in that it allows us to treat a group of agents like a single agent when it comes to assessing that group's
responsibility:
\begin{description}
\item[(IGC)] {\em Independence of Group Composition.}
  If $i,i'\in G$ and all occurrences of $i'$ are replaced in $\T$ by $i$, $\R(G)$ remains unchanged.
\end{description}
The final two in this category of axioms basically state that certain forms of luck should not influence responsibility assessments:
\begin{description}
\item[(FIU)] {\em Forward Independence of Unknowns.}
  If $v_d\sim v_d'\in V_i$ then $\R_f(v_d,\{i\}) = \R_f(v_d',\{i\})$ 
  (i.e., forward responsibility is the same in decision nodes the agent cannot distinguish).
\item[(BIL)] {\em Backward Independence of Luck.} 
  If $v_o,v_o'\in V_o$, $H(v_o)\cap V_d = H(v_o')\cap V_d = W$, and $C_{v''}(v_o)=C_{v''}(v_o')$ for all $v''\in W$,
  then $\R_b(v_o) = \R_b(v_o')$
  (i.e., backward responsibility is the same in outcome nodes that have the same choice history).
\end{description}
The combination of all the above axioms would allow us to restrict our interest to 
single-agent situations that have $I=G=\{i\}$, 
have only properly branching non-outcome nodes,
at most one ambiguity node and only as their root node, 
have no two consecutive probability nodes, no zero probabilities,
and no consecutive decision nodes.

\paragraph{Continuity, monotonicity, and other inequality axioms.}
The axioms in this category state how an RF may change when certain features of the situation change.
The first one disallows slight changes in probabilities to have large impacts on assessments: 
\begin{description}
\item[(PCont)] {\em Probability Continuity.}
  If $v\in V$ and for any $v_p\in V_p$, the probability distribution $p_{v_p}$ is varied continuously, 
  then $\R(v)$ does not change discontinuously in dependence on $p_{v_p}$.
\end{description}
The next four basically state that reduced agency or certain forms of ambiguity should not increase responsibility:
\begin{description}
\item[(CAM)] {\em Current Agency Monotonicity.}
  If we remove some action $a\in A_{v_d}$ and its branch $B(c_{v_d}(a))$ from the current decision node $v_d\in V_G$ and the latter is complete-information, then $\R_f(v_d)$ does not increase. 
\item[(PAM)] {\em Past Agency Monotonicity.}
  If we remove some non-taken action $a\in A_{v_d'}$, $C_{v'}(v_o)\neq a$, and its branch from a past decision node $v_d'\in V_G\cap H(v_o)$ and the latter is complete-information,
  then $\R_b(v_o)$ does not increase.
\item[(AMF)] {\em Ambiguity Monotonicity of Forward Responsibility.}
  If we remove the branch $B(v')$ of a possible successor $v'\in S_{v_a}$ of an ambiguity node $v_a\in V_a$ that does not contain $v$, $v\notin B(v_a)$, 
  then $\R_f(v)$ does not increase.
\end{description}

The last four relate responsibilities of groups to their subgroups, and backward to forward responsibility:
\begin{description}
\item[(GSM)] {\em Group Size Monotonicity.}
  If $G\subseteq G'$ then $\R(G)\le \R(G')$
  (i.e., larger groups have no less responsibility).
\item[(GSA)] {\em Group Subadditivity.}
  $\R(G + G') \le \R(G) + \R(G')$ for all $G,G'\subseteq I$.
\item[(GPA)] {\em Group Superadditivity.}
  $\R(G + G') \ge \R(G) + \R(G')$ for all $G,G'\subseteq I$.
\item[(GA)] {\em Group Additivity.}
  $\R(G + G') = \R(G) + \R(G')$ for all disjoint $G,G'\subseteq I$.
\item[(MBF)] {\em Maximal Backward Responsibility Bounds Forward Responsibility.} 
  For all $v,G$, there must be $\sigma\in\Sigma(v,G)$ and $v_o\in V_o^\sigma$ 
  so that $R_b(v_o,G) \ge \R_f(v,G)$
  (i.e., forward responsibility is bounded by potential backward responsibility).
\end{description}

\paragraph{Existence and special situation axioms.}
The first two axioms in this category require the existence of responsible groups
and responsibility-avoiding strategies.
\begin{description}
\item[(NRV)] {\em No Responsibility Voids.}
  If $V_a=V_p=\emptyset$ and $\eps\neq V_o$,
  then for each $v_o\in\eps$, 
  there is $G\subseteq I$ with $\R_b(v_o,G) > 0$.
\item[(NUR)] {\em No Unavoidable Backward Responsibility.} 
  For each $G\subseteq I$, there exists a strategy $\sigma\in\Sigma(v_0,G)$ 
  so that $\R_b(v_o,G) = 0$ for all $v_o\in V_o^\sigma$
  (i.e., $G$ must have a way of avoiding backward responsibility).
\end{description}
Finally, we consider a number of axioms which require certain values of $\R_b$ or $\R_f$ 
for the paradigmatic example situations discussed informally in the Introduction.
\begin{description}
\item[(Norm)] {\em Responsibility Degree Normalization.}
  With $\T$ and $\eps$ as depicted in Fig.\ \ref{fig:paradigmatic}(c) with $p=0$ and $q=1$, 
  $\R_f(v_1,\{i\})=\R_b(v_5,\{i\})=1$ and $\R_b(v_2,\{i\})=0$.
\item[(NWT)] {\em No Wishful Thinking.}
  With $\T$ and $\eps$ as depicted in Fig.\ \ref{fig:paradigmatic}(a), 
  $\R_f(v_2,\{i\})=1$.
\item[(NUT)] {\em No Unfounded Trust.}
  With $\T$ and $\eps$ as depicted in Fig.\ \ref{fig:paradigmatic}(a), 
  $\R_f(v_1,\{i\})=1$.
\item[(NFT)] {\em No Fearful Thinking.}
  With $\T$ and $\eps$ as depicted in Fig.\ \ref{fig:paradigmatic}(b), 
  $\R_f(v_1,\{i\})=1$.
\item[(NUD)] {\em No Unfounded Distrust.}
  With $\T$ and $\eps$ as depicted in Fig.\ \ref{fig:paradigmatic}(b), 
  $\R_f(v_2,\{i\})=1$.
\item[(UFR)] {\em Undivided Factual Responsibility.} 
  With $\T$ and $\eps$ as depicted in Fig.\ \ref{fig:paradigmatic}(a), 
  $\R_b(v_6,\{i\})=1$.
\item[(MFR)] {\em Multicausal Factual Responsibility.} 
  With $\T$ and $\eps$ as depicted in Fig.\ \ref{fig:paradigmatic}(b), 
  $\R_b(v_6,\{i\})=1$.
\item[(CFR)] {\em Counterfactual Responsibility.} 
  With $\T$ and $\eps$ as depicted in Fig.\ \ref{fig:paradigmatic}(a), 
  $\R_b(v_4,\{i\})=1$.
\item[(OPR)] {\em Ordered Probability Responsiveness.}
  If $\T$ and $\eps$ are as depicted in Fig.\ \ref{fig:paradigmatic}(c), 
  $p < q$, and we either increase $q$ or decrease $p$,
  then $\R_f(v_1,\{i\})$ and $\R_b(v_5,\{i\})$ strictly increase.
\item[(MAR)] {\em Multiple Aberration Responsiveness.}
  If $\T$ and $\eps$ are as depicted in Fig.\ \ref{fig:paradigmatic}(d) and $p>0$, 
  $\R_b(v_5,\{i\}) > \R_b(v_6,\{i\})$
  (i.e., an uncertain additional chance to avoid $\eps$ strictly increases backward responsibility if missed).
\end{description}
Note that (AMF) and (Norm) together imply (NWT) and (NFT) but not (NUT) or (NUD),
while (NWT) and (FIU) imply (NUT), and (NFT) and (FIU) imply (NUD). 
Also, given (FCS), (NWT) and (NFT) become equivalent, and (NUT) and (NUD) become equivalent.
Finally, (BIL), (IAT), and (UFR) together imply (CFR).

\subsection*{Proofs and further propositions}

\def\QED{{\em Q.E.D.}}

{\em Proof} of Proposition \ref{prop:main}.
Whenever the consequences of some change in $\T$ are discussed, quantities after the change are marked by $\hat~$.
Let $B_0(v)=B(v)\cap V_o$.
\begin{description}

\item[(IND)] Fig.\ \ref{fig:ind} is a counterexample for $\R^3$ and $\R^4$.
Compliance of $\R_b^0$ is straightforward.
Because of the canonical bijection between the strategy sets $\Sigma$ before and after the change,
at no node $v\neq v_d'$, $\gamma(v)$ or $\mu(v)$ change,
hence at no node $v\neq v_d$, $\R_f^1$ or $\R_f^2$ change.
Hence, if $C_{v_d}(v_o)\neq a$, no summand $\Delta\gamma$ or $\Delta\mu$ occurring in $\R_b^1(v_o)$ or $\R_b^2(v_o)$ changes.
If $C_{v_d}(v_o) = a$, $C_{v_d'}(v_o) = a'$ and $c_{v_d'}(a') = v''$, the only change is that the old 
$\Delta\gamma(v_d,a) + \Delta\gamma(v_d',a') = \gamma(c_{v_d}(a)) - \gamma(v_d) + \gamma(c_{v_d'}(a')) - \gamma(v_d') = \gamma(v'') - \gamma(v_d)$
is replaced by the new 
$\Delta\gamma(v_d,(a,a')) = \gamma(c_{v_d}(a,a')) - \gamma(v_d) = \gamma(v'') - \gamma(v_d)$,
which is the same value. The same holds for $\mu$ instead of $\gamma$.

\item[(IAT)] Fig.\ \ref{fig:iat} becomes a counterexample for $\R^0$ and $\R^1$ if one puts $\eps = \{v'\}$ and $v_o = v'$.
Because of the canonical bijection between the scenario sets $Z^\sim$ before and after the change
and since $\R^2$, $\R^3$, and $\R^4$ use $Z^\sim$ rather than just $Z$, the values of $\mu$ and $\omega$ do not change,
hence these varianty comply with (IAT).
(Note that $\R^1$ would also comply if we had used $Z^\sim$ rather than $Z$ in the definition of $\gamma$, 
and $\R^0$ would comply if we had used $B^\sim$ rather than $B$ in its definition.)

\item[(GSM)] Counterexamples for $\R^2$, $\R^3$, and $\R^4$ can be constructed easily.  

\item[(AMF)] If $v_a\notin B(v_d)$, none of the variants of $\R_f(v_d)$ are affected by the change. So assume $v_a\in B(v_d)$.
Since the change reduces the scenario set $Z^\sim(v_d)$ but does not alter any $\Delta_\ell(v,\zeta)$ for any remaining scenario $\zeta$, 
the value of the maximum $\R^3_f(v_d)$ cannot increase.

A counterexample for $R^1$, $R^2$ and $\R^4_f$ is the situation where the two available choices at some decision node $v_d$ lead to two ambiguity nodes $v_d$, $v_d'$,
each of which has two successors, one of which is in $\eps$ and the other not in $\eps$. 
Then both choices have guaranteed likelihood of 0, a minimax likelihood 1, and are equally risky, hence $\R^1_f(v_d)=\R^2_f(v_d)=\R^4_f(v_d)=0$. 
But after removing the successor of $v_d$ that is not in $\eps$,
the first choice has guaranteed likelihood 1, hence $\R^1_f(v_d)=1$ after the removal.
Likewise, after removing the successor of $v_d$ that is in $\eps$,
the first choice has minimax likelihood 0 and ceases to be risky, hence $\R^2_f(v_d)=\R^4_f(v_d)=1$ after the removal.

\item[(NRV)] Fig.\ \ref{fig:learn}(a) is a counterexample for $\R^2$ and $\R^4$.

\item[(NUR)] Fig.\ \ref{fig:learn}(a) is a counterexample for $\R^3$.

\end{description}
All other properties should be obvious from the definitions. \QED

~

\begin{proposition}
All variants $\R^0$, $\R^1$, $\R^2$, $\R^3$, $\R^4$ fulfill 
(Anon), (ACon), (OCon), (IST), (ICP), (INA), (INP), (IOA), (IGC), (BIL), and (UFR).

The variants $\R^1$, $\R^2$, $\R^3$, $\R^4$ also fulfill
(IZP), (PCont), (CAM), (Norm), (OPR), and (MAR),
while $\R^0$ fulfills neither.

The variants $\R_f^1$, $\R_f^2$, $\R_f^3$, $\R_f^4$ also fulfill
(NWT).

(FIU) and (NUT) are fulfilled by $\R_f^2$, $\R_f^3$, $\R_f^4$ but not $\R_f^1$.
\end{proposition}
{\em Proof.}
\begin{description}

\item[(IST)]
$\R^0$: If $S_v=\{v'\}$, $B_o(v)=B_o(v')$, so the change does not affect $\R^0$.

$\R^1$: There are obvious canonical bijections $F$ between the scenario sets $Z^\sim(v)$ and $\hat Z^\sim(v)$ 
and between the strategy sets $\Sigma(v)$, $\hat\Sigma(v)$ before and after the change.
Since $\Delta\gamma(v_d,a)$ is unaffected for $v_d\neq v$, while $\Delta\gamma(v_d,a) = 0$ and $\Delta\hat{\gamma}(v_d,a) = \Delta\gamma(v')$ if $v_d=v$, 
its sum along $H(v_o)\cap V_G$ (giving $\R_b^1(v_o)$) is unaffected,
and $\R_f^1(v_d)$ is unaffected for $v_d\neq v$, while $\R_f^1(v_d) = 0$ and $\hat\R_f^1(v_d) = \R_f^1(v')$ if $v_d=v$.

$\R^2$: As for $\R^1$, with $\Delta\mu(v_d,a)$ instead of $\Delta\gamma(v_d,a)$. 

$\R^3$, $\R^4$: As for $\R^1$, with $\rho(v_d,a)$ instead of $\Delta\gamma(v_d,a)$.

\item[(INA)]
$\R^0$: Note that $B_o(v)$ for any $v\neq v_a'$ remains unchanged when pulling $v_a'$ back into $v_a$.
Let $v_o\in V_o$, $v\in H(v_o)$, and $v_d=P(v)\in V_d$.
Since $v_d\in V_d$, $v_d\neq v_a, v_a'$ and thus $v\neq v_a'$. 
Hence $B_o(v)\subseteq\eps\not\supseteq B_o(v_d)$ after the change if and only if this is so before the change.

$\R^1$, $\R^2$: Since there is an obvious canonical bijection between the scenario sets before and after the change,
$\gamma(v)$ and $\mu(v)$ do not change for any $v\neq v_a'$, 
$\Delta\gamma(v_d,a)$ and $\Delta\mu(v_d,a)$ not for any $v_d$, and $H(v_o)\cap V_G$ not for any $v_o$, hence $\R^1$ and $\R^2$ are unaffected.

$\R^3$, $\R^4$: Let $F$ be the above-mentioned bijection between scenarios. 
Since $\hat\omega(v,F(\zeta))=\omega(v,\zeta)$ and $\Delta\hat{\ell}(v,F(\zeta))=\Delta\ell(v,\zeta)$ for all $v\neq v_a'$, 
also $\hat\rho(v_d,a)=\rho(v_d,a)$, so $\Delta\rho(v_d)$ and thus both $\R^3$ and $\R^4$ are unaffected.

\item[(INP)] This is completely analogous to (INA).

\item[(IZP), (PCont)]
The change discussed in (IZP) does not affect any $\ell(\eps|v'',\sigma,\zeta)$ for $v''\notin B(v')$, 
and that in (PCont) affects $\ell$ continuosly.
Since $\R^1$--$\R^4$ are based on $\ell$ alone and are continuous in $\ell$, the claim follows.
Minimal counterexamples for $\R^0$ are trivial to find.

\item[(CAM)] The change removes some strategies, hence $\gamma(v_d)$, $\mu(v_d)$, and $\omega(v_d,\zeta)$ can not decrease,
while $\gamma(c_{v_d}(a'))$, $\mu(c_{v_d}(a'))$, and $\omega(c_{v_d}(a')),\zeta)$ are unaffected for all $a'\neq a$,
so that $\Delta\gamma(v_d,a')$, $\Delta\mu(v_d,a')$, and $\Delta\omega(v_d,\zeta,a')$ cannot increase for any $a'\neq a$.
The variants of $\R_f$ are weakly monontonic functions of the latter.

\item[(FIU)] $\R^2$, $\R^3$, $\R^4$ are based on $Z^\sim$, while $\R^1$ uses $Z$ and thereby ignores information equivalence.
\end{description} 
All other properties should be obvious from the definitions. \QED

~

Regarding (PAM), we abstain from proving the following conjecture.
\begin{conjecture}
All variants $\R^0$, $\R^1$, $\R^2$, $\R^3$, $\R^4$ fulfill (PAM).
\end{conjecture}

\subsection*{Non-graded variant based on the NESS condition}

Here we finally sketch some BRF variant based on the idea of the NESS criterion \cite{braham2009degrees}, interpreting their notion of `event' in our context as a single decision taken by some agent, acknowledging the information sets of agents.

An {\em information set} for $i$ is a $\sim$-equivalence-class $y\subseteq V_i$. 
Let $Y_i$ be the set of all information sets for $i$
and $Y_G=\bigcup_{i\in G}Y_i$.
Then $Y=\bigcup_i Y_i$ is the partition of $V_d$ into $\sim$-equivalence classes.
For $v_d\in V_d$, let $y(v_d)$ be that $y\in Y$ with $v_d\in y$.
A {\em decision} is a pair $d=(y,a)$ with $y\in Y$ and $a\in A_{v_d}$ for all $v_d\in y$. 
For an outcome $v_o\in V_o$, let 
$$D(v_o) = \{ (y(v_d),C_{v_d}(v_o)) : v_d\in H(v_o)\cap V_d\}$$ 
be the set of all taken decisions that led to $v_o$.
For a set $D$ of decisions, let 
$$ V_o^D = \{ v_o\in V_o : D\subseteq D(v_o) \} $$
be the set of all outcomes that may occur if all the decisions in $D$ are actually taken. 
Note that $V_o^D$ is a weakly decreasing set function of $D$.

Let some $v_o\in\eps$ be fixed.

A subset $D\subseteq D(v_o)$ of the decisions that led to $v_o$ is called {\em sufficient} iff $V_o^D\subseteq \eps$.
Note that if $D$ is sufficient, so is every larger $D'\supseteq D$,
but there need not be any sufficient set since the whole $D(v_o)$ might not be sufficient if luck plays a role (e.g., in Fig.~\ref{fig:paradigmatic}(c)).

A decision $d$ is {\em necessary for the sufficiency} of $D$ iff $d\in D$, $D$ is sufficient, but $D-d$ is not sufficient.

A decision $d\in D(v_o)$ was a {\em NESS-cause} for $\eps$ if there is some $D\subseteq D(v_o)$ such that $d$ is necessary for the sufficiency of $D$.

A group $G$ is {\em NESS-responsible} for $\eps$, denoted $\R_b^N(i,v_o)=1$, iff they took some decision $(y,a)\in D(v_o)$,  $y\in Y_G$, that was a NESS-cause for $\eps$. 

It appears that the resulting BRF $\R_b^N$ probably fulfills the axioms (IGC), (IND), (IAT), (GSM), (NRV), (MFR)
but probably violates (CFR) and (IOA) (if the converted node belonged to a non-singleton information set).
Which of the other axioms it fulfills is beyond the scope of this paper.

\pagebreak[4]
\global\pdfpageattr\expandafter{\the\pdfpageattr/Rotate 90}
\begin{sidewaystable}
\footnotesize
\begin{tabular}{llllllllll
}\toprule
    \textbf{Method} 
        & $\bm{\R_f^1}$     & $\bm{\R_f^2}$     & $\bm{\R_f^3}$     & $\bm{\R_f^4}$ 
            & $\bm{\R_b^0}$
                & $\bm{\R_b^1}$     & $\bm{\R_b^2}$         & $\bm{\R_b^3}$                 & $\bm{\R_b^4}$ \\ \midrule
    two-option majority voting 
        & $1_{m > \N2}$     & $1_{m > \N2}$     & 1                 & 1 
            & $1_{u > \N2}$                 
                & $1_{u > \N2}$     & $1_{u > m - \N2 > 0}$ & $1_{u > \max(m - \N2, 0)}$    & $1_{u > \max(m - \N2, 0)}$ \\ \midrule
    multi-option simple majority 
        & $1_{m > \N2}$     & $1_{m > \N2}$     & 1                 & $\bm{1_{m\ge \N2-1}}$
            & $1_{u > \N2}$                 
                & $1_{u-a>m'}$      & $1_{a-u<m'<m}$       & $1_{a-u<\min(m',m-1)\atop\text{~or~}\bm{m<\N2-1}}$       & $1_{a-u<\min(m',m-1)}\atop\times 1_{m\ge \N2-1}$ \\ \midrule
    approval voting 
        & $1_{m > \N2}$     & $1_{m > \N2}$     & 1                 & $1$
            & $1_{u > \N2}$                 
                & $1_{u-a>m'}$      & $1_{a-u<m'<m}$       & $1_{a-u<\min(m',m-1)}$       & $1_{a-u<\min(m',m-1)}$ \\ \midrule 
    any single-round majoritarian
        & $1_{m > \N2}$     & $1_{m > \N2}$     & 1                 & 
            & $\le 1_{m > \N2}$         
                & $\le 1_{m > \N2}$ & $\le 1_{m > \N2}$   \\ \midrule
    random dictator 
        & $m/N$             & $m/N$             & $m/N$             & $m/N$ 
            & $\bm{1_{u=N}}$         
                & $u/N$             & $u/N$                 & $u/N$                         & $u/N$ \\ \midrule
    full consensus / random dictator 
        & $m/N$             & $m/N$             & 1                 & $m/N$ 
            & $1_{u=N}$         
                & $\bm{1_{a<m}} u/N$& $u/N$                 & $(u+m')/N$                   & $u/N$ \\   \midrule
    any single-round prop. power alloc. 
        & $m/N$             & $m/N$             &                   &
            & $\le 1_{m=N}$         
                & $\le m/N$         & $\le m/N$    \\ \midrule
    polling round before actual voting 
        & 0                 & 0                 & 0                 & 0
            & 0                 
                & 0                 & 0                     & 0                             & 0 \\ \midrule
    amendment / majority, round 1
        & $\bm 0$           & $\bm 0$           & 1                 & 1     
            & $0$
                & $0$               & $0$                   & $1_{a<m<\N2}$                 & $1_{a<m<\N2}$\\
    amendment / majority, round 2
        & $1_{m > \N2}$     & $1_{m > \N2}$     & 1                 & 1
            & $+ 1_{b < m-\N2}$         
                & $+ 1_{b < m-\N2}$   & $+ 1_{b<\N2<m}$       & $+ 1_{b<\min(\N2,m)}$         & $+ 1_{b<\min(\N2,m)}$\\ \midrule
    simple runoff, round 1
        & $1_{m > \N2}$     & $1_{m > \N2}$     & 1                 & 1      
            & $1_{a+b<\N2}$   
                & $1_{a+b<\N2}$     & $1_{a+b<\N2<m}$               & $1_{a + b < \max(m - \frac N 6, \N2)}$ & $1_{a + b < \max(m - \frac N 6, \N2)}$ \\
    simple runoff, round 2
        & $1_{m > \N2}$     & $1_{m > \N2}$     & 1                 & 1      
            & $+ 1_{u>\N2}$   
                & $+ 1_{u>\N2}$     & $+ 1_{u>m-\N2>0}$             & $+ 1_{u > \max(m - \N2, 0)}$    & $+ 1_{u > \max(m - \N2, 0)}$ \\ \midrule
    median voting on emissions
        & $1_{m > \N2}$     & $1_{m > \N2}$     & 1                 & 1      
            & $1_{m > \N2}\atop\times 1_{a_{m-(N-1)/2}\ge c_1}$   
                & $1_{m > \N2}\atop\times f(a_{m-(N-1)/2})$ & $1_{m > \N2}\atop\times f(a_{(N+1)/2})$ & $f(a_{\min(m,(N+1)/2)})$ & $f(a_{\min(m,(N+1)/2)})$ \\
    \bottomrule
\end{tabular}
\caption{\label{tbl:vals}
Degrees of responsibility of a voter group $G$ of size $m$ out of all $N$ voters
regarding the event that an ethically undesired option $U$ gets elected.
$u,a,b$ denote certain numbers of votes from $G$,
and $c_0,c_1,f$ certain parameters and functions,
depending on the method used (see text). 
We assume $N$ is odd, $m'=N-m$, and exclude the knife's edge cases where $|u-a| \in \{ m', m-1 \}$ here since they are more complicated.
$1_C$ is the indicator function of condition $C$, 1 if $C$ is true, 0 otherwise.
Details highlighted in boldface may be seen as problematic.
}
\end{sidewaystable}

\end{document}